\documentclass[twocolumn,pra,showpacs,superscriptaddress]{revtex4-1}%
\usepackage{amssymb}
\usepackage{amsmath}
\usepackage{amsfonts}
\usepackage{amssymb}
\usepackage{graphicx}
\usepackage{txfonts}%
\providecommand{\U}[1]{\protect\rule{.1in}{.1in}}

\begin{document}
\title{Improving quantum parameter estimation by monitoring quantum trajectories}
\author{Yao Ma}
\author{Mi Pang}
\affiliation{Department of Applied Physics, School of Sciences, Xi'an University of
Technology, Xi'an 710048, China}
\author{Libo Chen}
\affiliation{School of Science, Qingdao University of Technology, Qingdao 266033, China}
\author{Wen Yang}
\email{wenyang@csrc.ac.cn}
\affiliation{Beijing Computational Science Research Center, Beijing 100193, China}

\begin{abstract}
Quantum-enhanced parameter estimation has widespread applications in many
fields. An important issue is to protect the estimation precision against the
noise-induced decoherence. Here we develop a general theoretical framework for
improving the precision for estimating an arbitrary parameter by monitoring
the noise-induced quantum trajectorie (MQT) and establish its connections to
the purification-based approach to quantum parameter estimation. MQT can be
achieved in two ways: (i) Any quantum trajectories can be monitored by
directly monitoring the environment, which is experimentally challenging for
realistic noises; (ii) Certain quantum trajectories can also be monitored by
frequently measuring the quantum probe alone via ancilla-assisted encoding and
error detection. This establishes an interesting connection between MQT and
the full quantum error correction protocol. Application of MQT to estimate the
level splitting and decoherence rate of a spin-1/2 under typical decoherence
channels demonstrate that it can avoid the long-time exponential loss of the
estimation precision and, in special cases, recover the Heisenberg scaling.

\end{abstract}

\pacs{06.20.-f, 03.65.Yz, 42.50.Dv}
\maketitle

\section{Introduction}

The precise estimation of parameters characterizing physical processes
\cite{GiovannettiPRL2006,DegenRMP2017} has applications in many fields, such
as gravitational-wave detection \cite{CavesPRD1981,LIGONatPhys2013}, frequency
spectroscopy \cite{WinelandPRA1992,BollingerPRA1996}, magnetometry
\cite{BudkerNatPhys2007,RondinRPP2014}, optical phase estimation
\cite{DowlingContempPhys2008}, and atomic clocks \cite{BorregaardPRL2013}.
With classical probes, repeated measurements can be used to improve the
estimation precision according to the classical $1/\sqrt{N}$ scaling with
respect to the number $N$ of repetitions. With quantum probes, quantum
resources (such as entanglement) can be utilized to improve the estimation
beyond the classical scaling and even attain the fundamental Heisenberg $1/N$
scaling allowed by quantum mechanics, where $N$ is the number of probes used
in the estimation. However, the inevitable presence of environmental noises
decoheres the quantum probes \cite{YangRPP2017}, limits the available quantum
resources, and severely degrades the estimation precision. This poses a
critical challenge to the practical realization of quantum-enhanced parameter estimation.

To address this problem, several methods have been developed, such as
dynamical decoupling
\cite{MazeNature2008,TaylorNatPhys2008,ZhaoNatNano2011,CaiNJP2013,LiPRA2015}
(see Refs. \onlinecite{YangFP2011,DegenRMP2017} for a review), time
optimization \cite{MatsuzakiPRA2011,ChinPRL2012,ChavesPRL2013}, and quantum
error correction (QEC)
\cite{DuerPRL2014,KesslerPRL2014,ArradPRL2014,LuNC2015,HerreraMartiPRL2015,UndenPRL2016,BergmannPRA2016,MatsuzakiPRA2017}
or feedback control \cite{ZhengPRA2015,HiroseNature2016,LiuPRA2017a}. The idea
of dynamical decoupling is to apply pulsed
\cite{TaylorNatPhys2008,WatanabePRL2010,LaraouiNatCommun2013,TanPRA2013,ZhaoPRA2014a,
ZhaoPRA2014,LiPRA2015,BossPRL2016,MaPRA2016,MaPRA2016a} or continuous
\cite{CaiNJP2013,YanNatComm2013,LoretzPRL2013,LangPRX2015} control on the
quantum probe to reduce its coupling to the noise and hence prolong its
coherence time. It has achieved remarkable success in detecting alternating
signals \cite{KotlerNature2011,LangePRL2011}, noises
\cite{LangeScience2010,AlvarezPRL2011,MedfordPRL2012,BarGillNatCommun2012,
YanNatComm2013,LoretzPRL2013,MuhonenNatNano2014}, and other quantum objects
\cite{ZhaoNatNano2012,KolkowitzPRL2012,TaminiauPRL2012,LondonPRL2013,
StaudacherScience2013,MaminScience2013,ShiNatPhys2014}, but it is only
applicable to non-Markovian noises
\cite{LuPRA2010,BerradaPRA2013,SekatskiNJP2016,WangNJP2017}. The idea of time
optimization is to mitigate decoherence by shortening the evolution time of
the quantum probe. It can improve the\textit{ }scaling of the estimation
precision beyond the classical scaling, but requires vanishingly short
evolution time and large-scale entanglement. Note that many Markovian
environments only allow the classical $1/\sqrt{N}$ scaling even if the most
general scheme is employed
\cite{FujiwaraJPA2008,EscherNatPhys2011,DobrzanskiNatCommun2012,EscherPRL2012,JarzynaPRL2013,DemkowiczPRL2014,DemkowiczBook2015,DemkowiczPRX2017,ZhouNC2018}%
. In this case, using short-range correlated states, which can be modeled by
matrix product states \cite{JarzynaPRL2013}, already gives almost optimal
performance. The idea of QEC is to detect and then correct the noise-induced
erroneous evolution. This is a powerful method applicable to both Markovian
and non-Markovian noises
\cite{DuerPRL2014,KesslerPRL2014,ArradPRL2014,LuNC2015,HerreraMartiPRL2015,UndenPRL2016,BergmannPRA2016,MatsuzakiPRA2017}%
. For Hamiltonian parameter estimation, recent works
\cite{DemkowiczPRX2017,ZhouNC2018} show that when the unitary Hamiltonian
evolution can be distinguished from the noise-induced evolution, QEC can even
recover the ultimate Heisenberg scaling; otherwise only a constant-factor
improvement over the classical scaling is possible.

Very recently, an interesting method was proposed
\cite{CatanaPRA2014,AlbarelliNJP2017,AlbarelliQ2018} to improve the estimation
precision. The idea is to monitor the environment
\cite{GammelmarkPRA2013,GammelmarkPRL2014,GenoniPRA2017} continuously to
(fully or partially) extract the information that leaks into the environment.
For certain Markovian environment, this method can recover the Heisenberg
scaling \cite{CatanaPRA2014,AlbarelliNJP2017,AlbarelliQ2018}, but previous
studies focus on specific Markovian environments and measurements and usually
relies on Gaussian approximation or numerically solving the stochastic master
equations. Moreover, this method requires direct measurement on the
environment, which is very challenging for realistic noise processes.

In this work, we try to address the above problems. First, we develop a
general theoretical framework for improving the precision of parameter
estimation via continuous monitoring of a general (either Markovian or
non-Markovian) environment and further establish its connection to the
purification-based approach to quantum parameter estimation
\cite{FujiwaraJPA2008}, which has motivated many works that derive fundamental
bounds on the estimation precision
\cite{EscherNatPhys2011,DobrzanskiNatCommun2012,EscherPRL2012,JarzynaPRL2013,ChavesPRL2013,DemkowiczPRL2014,DemkowiczBook2015,YuanPRA2017,SekatskiQ2017,DemkowiczPRX2017,ZhouNC2018}%
. Second, for Markovian environment, we provide a superoperator approach to
determining the fundamental bounds on the estimation precision. This approach
corresponds to an exact integration of the stochastic master equation
\cite{AlbarelliNJP2017,AlbarelliQ2018}, and may provide exact analytical
expressions for some simple models. Third, we relax the conceptually simple
but experimentally challenging requirement of monitoring the environment to
the concept of monitoring the quantum trajectories (referred to as MQT for
brevity):\ any quantum trajectories can be monitored by monitoring the
environment, but \textit{certain} quantum trajectories can also be monitored
by frequently measuring the quantum probe (without monitoring the environment)
via ancilla-assisted encoding and error detection
\cite{DuerPRL2014,KesslerPRL2014,ArradPRL2014}, i.e., the first two steps of
QEC. This QEC-based MQT not only makes certain MQT experimentally feasible,
but also establishes an interesting connection between MQT and the full
QEC-based metrology
\cite{DuerPRL2014,KesslerPRL2014,ArradPRL2014,LuNC2015,HerreraMartiPRL2015,UndenPRL2016,BergmannPRA2016,MatsuzakiPRA2017}%
. The QEC-based MQT can be regarded as a QEC protocol without corrective
operations, so it is less powerful than QEC when perfect error correction is
available. Nevertheless, MQT itself provides an insight into how the
information leaks into the distinct quantum trajectories and how they are
recovered. Moreover, for certain models where corrective operations are not
necessary, the MQT becomes advantageous because it avoids faulty corrective
operations that may degrade the estimation precision significantly
\cite{KesslerPRL2014}. We apply this method to the estimation of the level
splitting $\omega$ and the decoherence rate $\gamma$ of a spin-1/2 under three
decoherence channels: spin relaxation, spin flip, and spin dephasing. We find
that it can significantly improve the precision for estimating $\omega$ under
the spin relaxation channel, avoid the exponential loss of the precision for
estimating $\gamma$ (estimating $\omega$) under all the decoherence channels
(under the spin flip channel) and even recover the Heisenberg scaling for
estimating $\omega$ under the spin dephasing channel.

This paper is organized as follows. In Sec. II, we give the general theory of
MQT. In Sec. III, we apply MQT to estimate the level splitting and decoherence
rate of a spin-1/2. In Sec. IV, we draw the conclusions.

\section{General idea and theory}

\begin{figure}[ptb]
\includegraphics[width=\columnwidth]{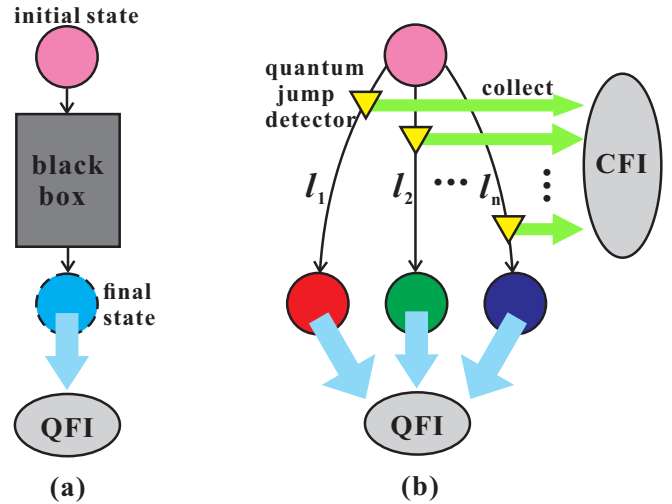}\caption{Quantum parameter
estimation by (a) conventional method and (b) monitoring quantum trajectories
(MQT):\ the former treats the noise-induced decoherence as a \textquotedblleft
black box\textquotedblright, while the latter gains access to the quantum
Fisher information (QFI)\ of every trajectory and the classical Fisher
information (CFI) contained in the timings of all the quantum jumps.}%
\label{G_SCHEME}%
\end{figure}

To estimate an unknown parameter $\theta$, the quantum probe starts from
certain initial state $\rho_{0}$ and then undergoes certain $\theta$-dependent
evolution for an interval $T$ into the final state $\rho(\theta)$, followed by
an optimal measurement on $\rho(\theta)$ to transfer all the information about
$\theta$ from the quantum probe into the measurement outcome. After repeating
the above procedures for $\nu\gg1$ times, we can use the $\nu$ measurement
outcomes to construct an optimal unbiased estimator to $\theta$, such as the
maximum likelihood estimator or the Bayesian estimator \cite{KayBook1993}. The
estimation precision for $\theta$ is determined by the quantum Cram\'{e}r-Rao
bound \cite{HelstromBook1976,BraunsteinPRL1994} as%
\begin{equation}
\delta\theta\equiv\frac{1}{\sqrt{\nu\mathcal{F}[\rho(\theta)]}}, \label{QCRB}%
\end{equation}
where $\mathcal{F}[\rho(\theta)]$ is the quantum Fisher information
(QFI)\ \cite{BraunsteinPRL1994} about $\theta$ provided by a single copy of
$\rho(\theta)$, while $\nu\mathcal{F}[\rho(\theta)]$ is the total QFI provided
by $\nu$ copies of $\rho(\theta)$. In Appendix A, we provide a detailed
introduction to all the relevant concepts, such as QFI, classical Fisher
information (CFI), optimal measurements, and optimal unbiased estimators.

\subsection{Non-unitary evolution and purification}

When the environment (or the quantum trajectories of the quantum probe) is not
monitored, the noise-induced decoherence during the $\theta$-dependent
evolution is a \textquotedblleft black box\textquotedblright\ [see Fig.
\ref{G_SCHEME}(a)], so our state of the knowledge about the quantum probe is
described by the non-selective density matrix $\bar{\rho}(t)$. The most
general non-unitary evolution of $\bar{\rho}(t)$ is described by a time-local
master equation \cite{KropfPRX2016,SmirnePRL2016}%
\[
\frac{d}{dt}\bar{\rho}(t)=\mathcal{L}(t)\bar{\rho}(t),
\]
where $\mathcal{L}(t)$ is a $\theta$-dependent Liouvillian, e.g.,
$\mathcal{L}(t)\bar{\rho}=-i[H(t),\bar{\rho}]$ in the absence of decoherence
or $\mathcal{L}(t)\bar{\rho}=-i[H(t),\bar{\rho}]+\sum_{a}\gamma_{a}%
(t)\mathcal{D}[c_{a}(t)]\bar{\rho}$ under a general decoherence channel, where
$\mathcal{D}[c]\rho\equiv c\rho c^{\dag}-\{c^{\dag}c,\rho\}/2$ describes the
decoherence in the Lindblad form, $\{c_{a}(t)\}$ are time-dependent quantum
jump operators, and $\{\gamma_{a}(t)\}$ are time-dependent decoherence rates.
The final state of the quantum probe is
\begin{equation}
\bar{\rho}(\theta)\equiv\mathcal{T}e^{\int_{0}^{T}\mathcal{L}(t)dt}\rho
_{0}\equiv\mathcal{N}_{\theta}(\rho_{0}), \label{RHO_NS_0}%
\end{equation}
where $\mathcal{T}$ is the time-ordering superoperator and $\mathcal{N}%
_{\theta}$ stands for the $\theta$-dependent non-unitary evolution -- the
quantum channel, which maps a $\theta$-independent initial state $\rho_{0}$ to
a $\theta$-dependent final state $\bar{\rho}(\theta)$. The non-unitary nature
of the quantum channel is manifested in the fact that the final state is mixed
even if the initial state is pure.

Recently, there were remarkable progress in establishing practical bounds on
the achievable estimation precision in the presence decoherence
\cite{FujiwaraJPA2008,EscherNatPhys2011,DobrzanskiNatCommun2012,EscherPRL2012,JarzynaPRL2013,ChavesPRL2013,DemkowiczPRL2014,DemkowiczBook2015,YuanPRA2017,SekatskiQ2017,DemkowiczPRX2017,ZhouNC2018}%
. The key idea is to purify the non-unitary quantum channel $\mathcal{N}%
_{\theta}$ of the quantum probe into a unitary evolution of an extended system
consisting of the quantum probe and an environment and then minimize the QFI
of the extended system \cite{FujiwaraJPA2008,EscherNatPhys2011}. In the
following, we use this purification formalism to establish a general theory of
MQT for an arbitrary environment.

When the environment causing the decoherence is included, the joint evolution
of the extended system (consisting of the quantum probe and the environment)
during $[0,T]$ is described by a unitary evolution operator $U_{\mathrm{ext}%
}(\theta)$ and the final state of the extended system is
\begin{equation}
\rho_{\mathrm{ext}}(\theta)\equiv U_{\mathrm{ext}}(\theta)(|E_{0}%
\rangle\langle E_{0}|\otimes\rho_{0})U_{\mathrm{ext}}^{\dagger}(\theta),
\label{RHO_EXT}%
\end{equation}
where $|E_{0}\rangle$ is the $\theta$-independent initial state of the
environment. Therefore, including the environment purifies the non-unitary
quantum channel $\mathcal{N}_{\theta}$ of the quantum probe into a unitary
evolution of the extended system, which maps a pure initial state
$|E_{0}\rangle\otimes|\psi_{0}\rangle$ into a pure final state
$U_{\mathrm{ext}}(\theta)|E_{0}\rangle\otimes|\psi_{0}\rangle$. Tracing out
the environmental degree of freedom in an arbitrary ortho-normal complete,
$\theta$-independent basis $\{|E_{l}\rangle\}$ gives the reduced density
matrix of the quantum probe:%
\begin{equation}
\bar{\rho}(\theta)=\operatorname*{Tr}\nolimits_{E}\rho_{\mathrm{ext}}%
(\theta)=\sum_{l}\Pi_{l}(\theta)\rho_{0}\Pi_{l}^{\dagger}(\theta)\equiv
\sum_{l}\tilde{\rho}_{l}(\theta), \label{RHO_NS1}%
\end{equation}
where
\begin{equation}
\Pi_{l}(\theta)\equiv\langle E_{l}|U_{\mathrm{ext}}(\theta)|E_{0}%
\rangle\label{PL}%
\end{equation}
are Kraus operators acting on the quantum probe. Equation (\ref{RHO_NS1})
gives a representation of the non-unitary quantum channel $\mathcal{N}%
_{\theta}$ in terms of a set of Kraus operators $\{\Pi_{l}(\theta)\}$
\cite{FujiwaraJPA2008}, or equivalently a representation of $\bar{\rho}%
(\theta)=\mathcal{N}_{\theta}(\rho_{0})$ [see Eq. (\ref{RHO_NS_0})] in terms
of a set of quantum trajectories $\tilde{\rho}_{l}(\theta)\equiv\Pi_{l}%
(\theta)\rho_{0}\Pi_{l}^{\dagger}(\theta)$, which occurs with a probability
$P_{l}(\theta)\equiv\operatorname*{Tr}\tilde{\rho}_{l}(\theta)$. The
completeness of the environmental basis $\sum_{l}|E_{l}\rangle\langle
E_{l}|=1$ leads to the completeness of the Kraus operators: $\sum_{l}\Pi
_{l}^{\dagger}(\theta)\Pi_{l}(\theta)=1$ and hence the normalization
$\operatorname*{Tr}\bar{\rho}(\theta)=\sum_{l}P_{l}(\theta)=1$. For a pure
initial state $\rho_{0}=|\psi_{0}\rangle\langle\psi_{0}|$ of the quantum
probe, the final state of the extended system is
\begin{equation}
|\Psi_{\mathrm{ext}}(\theta)\rangle=U_{\mathrm{ext}}(\theta)|E_{0}%
\rangle\otimes|\psi_{0}\rangle=\sum_{l}|E_{l}\rangle\otimes\Pi_{l}%
(\theta)|\psi_{0}\rangle, \label{PHI_SE}%
\end{equation}
where $\Pi_{l}(\theta)|\psi_{0}\rangle$ is a pure-state quantum trajectory of
the quantum probe.

Replacing $U_{\mathrm{ext}}(\theta)$ by $u_{E}(\theta)U_{\mathrm{ext}}%
(\theta)$ with $u_{E}(\theta)$ being an arbitrary unitary operator acting on
the environment leaves the quantum channel $\mathcal{N}_{\theta}$ and hence
the final state $\bar{\rho}(\theta)=\mathcal{N}_{\theta}(\rho_{0})$ of the
quantum probe invariant, but changes $\Pi_{l}(\theta)$ to
\[
\pi_{l}(\theta)\equiv\langle E_{l}|u_{E}(\theta)U_{\mathrm{ext}}(\theta
)|E_{0}\rangle=\sum_{l^{\prime}}u_{ll^{\prime}}(\theta)\Pi_{l^{\prime}}%
(\theta),
\]
where $u_{ll^{\prime}}(\theta)\equiv\langle E_{l}|u_{E}(\theta)|E_{l^{\prime}%
}\rangle$ is a unitary matrix, so it gives a different representation of the
non-unitary quantum channel $\mathcal{N}_{\theta}$ in terms of a different set
of Kraus operators $\{\pi_{l}(\theta)\}$, or equivalently, a representation of
$\bar{\rho}(\theta)=\mathcal{N}_{\theta}(\rho_{0})$ in terms of a different
set of quantum trajectories:\ $\bar{\rho}(\theta)=\sum_{l}\pi_{l}(\theta
)\rho_{0}\pi_{l}^{\dagger}(\theta)$. Therefore, the purification (and hence
representation) of the non-unitary quantum channel $\mathcal{N}_{\theta}$ is
not unique:\ given a purification $U_{\mathrm{ext}}(\theta)$ and hence a
representation $\{\Pi_{l}(\theta)\equiv\langle E_{l}|U_{\mathrm{ext}}%
(\theta)|E_{0}\rangle\}$ for $\mathcal{N}_{\theta}$, exhausting all possible
unitaries $u_{E}(\theta)$ exhausts all possible unitary purifications
$u_{E}(\theta)U_{\mathrm{ext}}(\theta)$ and hence all possible Kraus operator
representations $\{\pi_{l}(\theta)\equiv\sum_{l^{\prime}}u_{ll^{\prime}%
}(\theta)\Pi_{l^{\prime}}(\theta)\}$\ of $\mathcal{N}_{\theta}$.\ Physically,
this means that there are an infinite number of different environments that
lead to the same reduced evolution $\mathcal{N}_{\theta}$ of the quantum
probe. Next we consider a hierarchy of constraints on our ability to measure
the joint system and derive an hierarchy of inequalities for the precision for
estimating $\theta$. In the following, we omit the dependences of various
quantities on $\theta$ for brevity.

\subsection{An hierarchy of estimation precision}

Here we consider a fixed environment that purifies the non-unitary evolution
$\mathcal{N}$ of the quantum probe into a unitary evolution $U_{\mathrm{ext}}$
of the extended system consisting of the quantum probe and the environment.
Correspondingly, the final state $\bar{\rho}=\mathcal{N}(\rho_{0})$ of the
quantum probe is purified into $\rho_{\mathrm{ext}}$ in Eq. (\ref{RHO_EXT})
for the extended system.

First, when arbitrary joint measurements on the extended system are available,
we can make an optimal joint measurement (see Appendix A) on the extended
system to extract all the QFI $\mathcal{F}[\rho_{\mathrm{ext}}]$ in the final
state $\rho_{\mathrm{ext}}$, so the fundamental precision follows from Eq.
(\ref{QCRB}) as
\begin{equation}
\delta\theta_{\mathrm{ext}}\equiv\frac{1}{\sqrt{\nu\mathcal{F}[\rho
_{\mathrm{ext}}]}}. \label{DW_SE}%
\end{equation}
This fundamental estimation precision was considered in Refs.
\cite{GammelmarkPRL2014,AlbarelliNJP2017,AlbarelliQ2018} for the special case
of time-homogeneous Markovian quantum channel, as described by a
time-homogeneous master equation.

Second, when arbitrary joint measurements are not available, but arbitrary
separate measurements on the quantum probe and the environment are available,
we can utilize different measurements on the environment to unravel $\bar
{\rho}$ into different sets of quantum trajectories [see Fig. \ref{G_SCHEME}%
(b)]. Specifically, a projective measurement on the environment in an
arbitrary ortho-normal complete, $\theta$-independent basis $\{|E_{l}%
\rangle\}$ has a probability $P_{l}\equiv\operatorname*{Tr}\tilde{\rho}_{l}$
to yield an outcome $|E_{l}\rangle$ and the occurrence of this outcome
collapses the quantum probe into the corresponding quantum trajectory
$\tilde{\rho}_{l}\equiv\Pi_{l}\rho_{0}\Pi_{l}^{\dagger}$ [with $\Pi_{l}$ given
by Eq. (\ref{PL})], which can be normalized as $\rho_{l}=\tilde{\rho}%
_{l}/P_{l}$. The average amount of information in the measurement outcome is
quantified by the CFI
\begin{equation}
F[\{P_{l}\}]=\sum_{l}\frac{(\partial_{\theta}P_{l})^{2}}{P_{l}},
\label{CFI_TRAJ}%
\end{equation}
while the average amount of information in the quantum trajectory $\rho_{l}$
is quantified by the QFI $\mathcal{F}[\rho_{l}]$. The latter can be fully
extracted by an optimal measurement on the quantum probe (see Appendix A).
Therefore, the average amount of information extracted from a measurement on
the environment in the basis $\{|E_{l}\rangle\}$ and an optimal measurement on
the quantum probe is
\begin{equation}
\mathbb{F}\equiv F[\{P_{l}\}]+\sum_{l}P_{l}\mathcal{F}[\rho_{l}], \label{FMQT}%
\end{equation}
which coincides with the QFI\ $\mathcal{F}[\rho_{\mathrm{ext}|\{E_{l}\}}]$ in
the joint state
\begin{equation}
\rho_{\mathrm{ext}|\{E_{l}\}}\equiv\sum_{l}|E_{l}\rangle\langle E_{l}%
|\otimes\Pi_{l}\rho_{0}\Pi_{l}^{\dagger}=\sum_{l}P_{l}|E_{l}\rangle\langle
E_{l}|\otimes\rho_{l} \label{RHO_POST}%
\end{equation}
after measuring the environment in the basis $\{|E_{l}\rangle\}$. The joint
state before the measurement Eq. (\ref{RHO_EXT}) can be written as
\[
\rho_{\mathrm{ext}}\equiv\sum_{ll^{\prime}}|E_{l}\rangle\langle E_{l^{\prime}%
}|\otimes\Pi_{l}\rho_{0}\Pi_{l^{\prime}}^{\dagger},
\]
so the measurement on the environment removes all off-diagonal coherences in
the measurement basis $\{|E_{l}\rangle\}$. After repeating this procedure for
$\nu\gg1$ times, we can use the $\nu$ outcomes from the measurements on the
environment and the $\nu$ outcomes from the optimal measurements on the
quantum probe to construct an optimal unbiased estimator to $\theta$ (see
Appendix A). The fundamental estimation precision of this MQT method follows
from Eq. (\ref{QCRB}) as%
\begin{equation}
\delta\theta_{\mathrm{MQT}}\equiv\frac{1}{\sqrt{\nu\mathbb{F}}}.
\label{DW_MQT}%
\end{equation}
In a previous work, Albarelli \textit{et al}. \cite{AlbarelliNJP2017}
considered homodyne measurement on a Markovian bosonic environment (leading to
time-homogeneous Markovian dynamics) and arrived at Eq. (\ref{DW_MQT}) for
this specific model through straightforward (but somewhat tedious) derivation
with the assistance of both the classical Cram\'{e}r-Rao bound and the quantum
Cram\'{e}r-Rao bound. Here our analysis shows that:\ (i) Eq. (\ref{DW_MQT}) is
valid for general non-unitary dynamics and general (projective) measurements
on the environment; (ii) Eq. (\ref{DW_MQT}) follows directly from the quantum
Cram\'{e}r-Rao bound [Eq. (\ref{QCRB})]. A similar analysis has been used to
discuss quantum parameter estimation with post-selection \cite{CombesPRA2014}.

Third, if only the quantum probe can be measured, then we can use an optimal
measurement (see Appendix A) on the quantum probe to extract all the
QFI\ $\mathcal{F}[\bar{\rho}]$ in $\bar{\rho}$, so the estimation precision
follows from Eq. (\ref{QCRB}) as
\begin{equation}
\delta\bar{\theta}\equiv\frac{1}{\sqrt{\nu\mathcal{F}[\bar{\rho}]}}.
\label{DW_NS}%
\end{equation}

Since the evolution $\rho_{\mathrm{ext}}\rightarrow\rho_{\mathrm{ext}%
|\{E_{l}\}}$ and $\rho_{\mathrm{ext}|\{E_{l}\}}\rightarrow\bar{\rho
}=\operatorname*{Tr}_{E}\rho_{\mathrm{ext}|\{E_{l}\}}$ are both non-unitary,
while any $\theta$-independent quantum operation cannot increase the QFI
\cite{FerriePRA2014}, we have%
\begin{equation}
\mathcal{F}[\rho_{\mathrm{ext}}]\geq\mathbb{F}=\mathcal{F}[\rho_{\mathrm{ext}%
|\{E_{l}\}}]\geq\mathcal{F}[\bar{\rho}] \label{INEQ}%
\end{equation}
and hence%
\[
\delta\theta_{\mathrm{ext}}\leq\delta\theta_{\mathrm{MQT}}\leq\delta
\bar{\theta}.
\]
In the above, $\mathcal{F}[\bar{\rho}]$ ($\mathcal{F}[\rho_{\mathrm{ext}}%
]$)\ is uniquely determined by the quantum state $\bar{\rho}$ ($\rho
_{\mathrm{ext}}$), while $\rho_{\mathrm{ext}|\{E_{l}\}}$ and hence
$\mathbb{F}$ still depend on the measurement on the environment. Optimal MQT
requires choosing an optimal measurement basis $\{|E_{l}\rangle\}$ to maximize
$\mathbb{F}$. The second inequality, i.e., $\mathbb{F}\geq\mathcal{F}%
[\bar{\rho}]$ is just the extended convexity of the QFI
\cite{AlipourPRA2015,NgPRA2016}, so our analysis not only provides a
physically intuitive proof for the extended convexity of the QFI, but also
identifies the physical meaning of $\mathbb{F}$ as the QFI$\ \mathcal{F}%
[\rho_{\mathrm{ext}|\{E_{l}\}}]$ in the post-measurement state $\rho
_{\mathrm{ext}|\{E_{l}\}}$ \cite{CombesPRA2014}.

During the first-round revision of this manuscript after submission, we became
aware of a very recent work by Albarelli \textit{et al}. \cite{AlbarelliQ2018}%
, which gives a similar equation as Eq. (\ref{INEQ}) for the special case of
photon-counting and homodyne measurement on a Markovian bosonic environment.
They further conjectured that $\mathbb{F}$ is a non-decreasing function of the
measurement efficiency. Here our general formalism allows a simple
generalization: since an imperfect measurement can be regarded as a perfect
measurement followed by a non-unitary quantum operation, while any $\theta
$-independent quantum operation cannot increase the QFI \cite{FerriePRA2014},
$\mathcal{F}[\rho_{\mathrm{ext}|\{E_{l}\}}]$ and hence $\mathbb{F}$ is a
non-decreasing function of the measurement efficiency on the environment for
\textit{any} environment.

\subsection{Connection to purification-based QFI bounds}

Another advantage of our general formalism is that it provides an interesting
connection between the MQT approach and the minimization over purification
(MOP) technique in quantum parameter estimation \cite{FujiwaraJPA2008}, which
has motivated many works that derive fundamental bounds on the estimation
precision
\cite{EscherNatPhys2011,DobrzanskiNatCommun2012,EscherPRL2012,JarzynaPRL2013,ChavesPRL2013,DemkowiczPRL2014,DemkowiczBook2015,YuanPRA2017,SekatskiQ2017,DemkowiczPRX2017,ZhouNC2018}%
. In the context of MOP, the aim is to find the maximum of the QFI
$\mathcal{F}[\bar{\rho}]$ in the non-selective final state $\bar{\rho
}=\mathcal{N}(\rho_{0})$ by optimizing the initial state $\rho_{0}$. Due to
the convexity of the QFI, the maximum of $\mathcal{F}[\bar{\rho}]$ is always
attained by pure initial states, so it suffices to consider $\rho_{0}%
=|\psi_{0}\rangle\langle\psi_{0}|$. Even in this case, the final state
$\bar{\rho}$ is still mixed, so calculating $\mathcal{F}[\bar{\rho}]$ requires
diagonalizing $\bar{\rho}$, which becomes tedious when the Hilbert space of
the quantum probe is large. By contrast, for a $\theta$-dependent pure state
$|\psi\rangle$, its QFI\ can be easily evaluated by the formula
\cite{HelstromBook1976,BraunsteinPRL1994} $\mathcal{F}[\psi]=4(\langle
\partial_{\theta}\psi|\partial_{\theta}\psi\rangle-\langle\psi|i\partial
_{\theta}\psi\rangle^{2})$ (see Appendix A). Interestingly, Escher \textit{et
al}. \cite{EscherNatPhys2011} proves a purification-based definition of the
QFI:%
\begin{equation}
\mathcal{F}[\bar{\rho}]=\min_{\Psi_{\mathrm{ext}}}\mathcal{F}[\Psi
_{\mathrm{ext}}]=\min_{\{\Pi_{l}\}}4(\langle\psi_{0}\left\vert \alpha
\right\vert \psi_{0}\rangle-\langle\psi_{0}\left\vert \beta\right\vert
\psi_{0}\rangle^{2}), \label{QFI_DEF2}%
\end{equation}
where $\alpha\equiv\sum_{l}(\partial_{\theta}\Pi_{l}^{\dagger})(\partial
_{\theta}\Pi_{l})$ and $\beta\equiv i\sum_{l}\Pi_{l}^{\dagger}\partial
_{\theta}\Pi_{l}$ are Hermitian operators acting on the quantum probe and the
minimization runs over all possible purifications $|\Psi_{\mathrm{ext}}%
\rangle$ [see Eq. (\ref{PHI_SE})] of $\bar{\rho}$ or equivalently all possible
Kraus operator representations of $\mathcal{N}$. Independently, Fujiwara and
Imai \cite{FujiwaraJPA2008} proves
\begin{equation}
\mathcal{F}[\bar{\rho}]=\min_{\Psi_{\mathrm{ext}}}4\langle\partial_{\theta
}\Psi_{\mathrm{ext}}|\partial_{\theta}\Psi_{\mathrm{ext}}\rangle=\min
_{\{\Pi_{l}\}}4\langle\psi_{0}\left\vert \alpha\right\vert \psi_{0}\rangle.
\label{QFI_DEF1}%
\end{equation}
These two definitions are equivalent because the purification that saturates
Eq. (\ref{QFI_DEF1}) obeys $\langle\Psi_{\mathrm{ext}}|i\partial_{\theta}%
\Psi_{\mathrm{ext}}\rangle=\langle\psi_{0}\left\vert \beta\right\vert \psi
_{0}\rangle=0$ \cite{FujiwaraJPA2008,DemkowiczBook2015}. An upper bound for
the maximal QFI $\max_{\psi_{0}}\mathcal{F}[\bar{\rho}]$ can be obtained by
exchanging $\max_{\psi_{0}}$ and $\min_{\{\Pi_{l}\}}$
\cite{DobrzanskiNatCommun2012,KolodynskiNJP2013,DemkowiczPRL2014,DemkowiczBook2015,SekatskiQ2017,DemkowiczPRX2017,ZhouNC2018}%
:%
\[
\max_{\psi_{0}}\mathcal{F}[\bar{\rho}]\leq4\min_{\{\Pi_{l}\mathbf{\}}%
}\left\Vert \alpha\right\Vert ,
\]
where $\left\Vert A\right\Vert $ is the maximal eigenvalue of $\sqrt
{A^{\dagger}A}$. When in addition to the quantum probe, access to some
ancillas is available, this upper bound is attainable; otherwise this upper
bound is not necessarily attainable \cite{FujiwaraJPA2008,KolodynskiNJP2013}.

To make connection to the MQT method, we notice that when $\rho_{0}=|\psi
_{0}\rangle\langle\psi_{0}|$, the quantum trajectories are pure states:
$\tilde{\rho}_{l}=|\tilde{\psi}_{l}\rangle\langle\tilde{\psi}_{l}|=P_{l}%
|\psi_{l}\rangle\langle\psi_{l}|$, where $|\tilde{\psi}_{l}\rangle\equiv
\Pi_{l}|\psi_{0}\rangle$ is the un-normalized trajectory, $P_{l}=\langle
\tilde{\psi}_{l}|\tilde{\psi}_{l}\rangle$ is the occurrence probability, and
$|\psi_{l}\rangle\equiv|\tilde{\psi}_{l}\rangle/\sqrt{P_{l}}$ is the
normalized trajectory. The QFI $\mathcal{F}[\Psi_{\mathrm{ext}}]$ in the final
state $\rho_{\mathrm{ext}}=|\Psi_{\mathrm{ext}}\rangle\langle\Psi
_{\mathrm{ext}}|$ [see Eq. (\ref{PHI_SE})] of the extended system is%
\begin{align*}
\mathcal{F}[\Psi_{\mathrm{ext}}]  &  =4(\langle\psi_{0}\left\vert
\alpha\right\vert \psi_{0}\rangle-\langle\psi_{0}\left\vert \beta\right\vert
\psi_{0}\rangle^{2})\\
&  =F[\{P_{l}\}]+4\left[  \sum_{l}P_{l}\langle\partial_{\theta}\psi
_{l}|\partial_{\theta}\psi_{l}\rangle-\left(  \sum_{l}P_{l}\langle\psi
_{l}|i\partial_{\theta}\psi_{l}\rangle\right)  ^{2}\right]  .
\end{align*}
For comparison, the QFI in the post-measurement state $\rho_{\mathrm{ext}%
|\{E_{l}\}}=\sum_{l}P_{l}|E_{l}\rangle\langle E_{l}|\otimes|\psi_{l}%
\rangle\langle\psi_{l}|$ of the extended system is%
\[
\mathbb{F}=\mathcal{F}[\rho_{\mathrm{ext}|\{E_{l}\}}]=F[\{P_{l}\}]+\sum
_{l}P_{l}\mathcal{F}[\psi_{l}],
\]
where $\mathcal{F}[\psi_{l}]$ is the QFI in the normalized trajectory
$|\psi_{l}\rangle$. As discussed in the previous subsection, the inequality
$\mathcal{F}[\Psi_{\mathrm{ext}}]\geq\mathcal{F}[\rho_{\mathrm{ext}|\{E_{l}%
\}}]$ [see Eq. (\ref{INEQ})] follows from the simple fact that $\rho
_{\mathrm{ext}|\{E_{l}\}}$ is obtained from $|\Psi_{\mathrm{ext}}\rangle$ by a
non-unitary operation, which cannot increase the QFI \cite{FerriePRA2014}.
Alternatively, we rewrite their difference as%
\[
\mathcal{F}[\Psi_{\mathrm{ext}}]-\mathbb{F}=4(|\mathbf{u}|^{2}|\mathbf{v}%
|^{2}-|\mathbf{u}\cdot\mathbf{v}|^{2}),
\]
where $\mathbf{u}$ and $\mathbf{v}$ are two real vectors: $(\mathbf{u}%
)_{l}=\sqrt{P_{l}}$ and $(\mathbf{v})_{l}=\sqrt{P_{l}}\langle\psi
_{l}|i\partial_{\theta}\psi_{l}\rangle$. Therefore, the inequality
$\mathcal{F}[\Psi_{\mathrm{ext}}]\geq\mathbb{F}$ simply follows from the
Cauchy-Schwarz inequality. This inequality is saturated if and only if
$\mathbf{u}\propto\mathbf{v}$, i.e., when
\begin{equation}
\langle\psi_{l}|\partial_{\theta}\psi_{l}\rangle=\lambda(\theta), \label{CD}%
\end{equation}
where $\lambda(\theta)$ is an arbitrary $l$-independent constant.

Now we discuss the connection and distinction between the MOP technique and
the MQT method. In the context of MOP, the ultimate goal is to derive the
tightest upper bound for the maximal QFI\ $\max_{\psi_{0}}\mathcal{F}%
[\bar{\rho}]$, which characterizes the fundamental precision for estimating
the parameter $\theta$ of a given quantum channel $\mathcal{N}$
\textit{without} any access to the environment. As a result, the environment,
the purification $|\Psi_{\mathrm{ext}}\rangle$, and the QFI $\mathcal{F}%
[\Psi_{\mathrm{ext}}]$ are not physical objects but instead mathematical tools
for converting the direct evaluation of the mixed-state QFI $\mathcal{F}%
[\bar{\rho}]$ to a minimization problem:\ $\mathcal{F}[\bar{\rho}]=\min
_{\Psi_{\mathrm{ext}}}\mathcal{F}[\Psi_{\mathrm{ext}}]$. The key physics is
that the quantum channel $\mathcal{N}$ and hence the final state $\bar{\rho
}=\mathcal{N}(|\psi_{0}\rangle\langle\psi_{0}|)$ can be generated by an
infinite number of fictitious environments. Each distinct environment
corresponds to a distinct joint unitary evolution $U_{\mathrm{ext}}$ and hence
a distinct Kraus operator representation $\{\Pi_{l}\}$ [see Eq. (\ref{PL})] of
$\mathcal{N}$ and a distinct purification $|\Psi_{\mathrm{ext}}\rangle$ [see
Eq. (\ref{PHI_SE})] of $\bar{\rho}$. The purification-based definitions of the
QFI [Eq. (\ref{QFI_DEF2}) or (\ref{QFI_DEF1})] dictates: (i)\ $\mathcal{F}%
[\Psi_{\mathrm{ext}}]\geq\mathcal{F}[\bar{\rho}]$, i.e., including the
environment never decreases the QFI; (ii)\ there exists QFI-preserving
environments for which the joint state $|\Psi_{\mathrm{ext}}\rangle$ contains
the same QFI as the reduced state: $\mathcal{F}[\Psi_{\mathrm{ext}%
}]=\mathcal{F}[\bar{\rho}]$.

By contrast, in the context of the MQT method, we assume that we have access
to the physical environment that is coupled to the quantum probe. In this
case, the quantum probe and the physical environment undergoes physical
unitary evolution $U_{\mathrm{ext}}$ as determined by their physical
Hamiltonians and mutual couplings, so we no longer have any degree of freedom
to choose the environment. Here the joint state $|\Psi_{\mathrm{ext}}\rangle$
and its QFI$\ \mathcal{F}[\Psi_{\mathrm{ext}}]$ are completely determined by
the initial state $|\psi_{0}\rangle$ of the quantum probe, while
$\mathbb{F}=\mathcal{F}[\rho_{\mathrm{ext}|\{E_{l}\}}]$ also depends on the
basis $\{|E_{l}\rangle\}$ of the measurement on the environment. The ultimate
goal is to optimize the initial state $|\psi_{0}\rangle$ and the measurement
basis $\{|E_{l}\rangle\}$ for maximal $\mathbb{F}$, e.g., if we can find a
suitable basis that satisfies Eq. (\ref{CD}), then $\mathbb{F}$ attains its
maximum $\mathcal{F}[\Psi_{\mathrm{ext}}]$. Interestingly, the
purification-based definition of the QFI suggests that when the physical
environment happens to be QFI-preserving, i.e., $\mathcal{F}[\Psi
_{\mathrm{ext}}]=\mathcal{F}[\bar{\rho}]$, then Eq. (\ref{INEQ}) dictates
$\mathcal{F}[\Psi_{\mathrm{ext}}]=\mathbb{F}=\mathcal{F}[\bar{\rho}]$ and
hence $\delta\theta_{\mathrm{ext}}=\delta\theta_{\mathrm{MQT}}=\delta
\bar{\theta}$, i.e., including the environment provides no advantage in
improving the estimation precision.

\subsection{Superoperator approach for Markovian dynamics}

\label{SEC_SUPER}

Here we consider homogeneous Markovian quantum channel $\mathcal{N}%
=e^{\mathcal{L}\newline T}$ described by a time-independent Liouvillian
$\mathcal{L}$, e.g., $\mathcal{L}\bar{\rho}=-i[H,\bar{\rho}]$ in the absence
of decoherence or $\mathcal{L}\bar{\rho}=-i[H,\bar{\rho}]+\sum_{a}\gamma
_{a}\mathcal{D}[c_{a}]\bar{\rho}$ in the presence of decoherence. With all
detectable quantum jumps
\cite{PlenioRMP1998,GleyzesNature2007,PiiloPRL2008,YuPRL2008,VamivakasNature2010,
DelteilPRL2014,CampagnePRX2016,NaghilooNatComm2016} denoted by the
superoperator $\mathcal{J}$ (see the next subsection for an example),
$\mathcal{L}$ becomes the sum of $\mathcal{J}$ and $\mathcal{L}_{0}%
\equiv\mathcal{L}-\mathcal{J}$ and the non-selective final state of the
quantum probe unravels into all possible quantum trajectories [cf. Eq.
(\ref{RHO_NS1})]:
\begin{equation}
\bar{\rho}=\tilde{\rho}_{\oslash}+\int_{0}^{T}dt_{1}\tilde{\rho}_{t_{1}}%
+\int_{0}^{T}dt_{2}\int_{0}^{t_{2}}dt_{1}\tilde{\rho}_{t_{1}t_{2}}+\cdots,
\label{RHO_NS}%
\end{equation}
where
\[
\tilde{\rho}_{\oslash}\equiv e^{\mathcal{L}_{0}T}\rho_{0}%
\]
is the\ trajectory with no quantum jump,%
\[
\tilde{\rho}_{t_{1}}\equiv e^{\mathcal{L}_{0}(T-t_{1})}\mathcal{J}%
e^{\mathcal{L}_{0}t_{1}}\rho_{0}%
\]
is the trajectory with one exclusive quantum jump at $t_{1}$,
\[
\tilde{\rho}_{t_{1}t_{2}}\equiv e^{\mathcal{L}_{0}(T-t_{2})}\mathcal{J}%
e^{\mathcal{L}_{0}(t_{2}-t_{1})}\mathcal{J}e^{\mathcal{L}_{0}t_{1}}\rho_{0}%
\]
is the trajectory with two exclusive quantum jumps at $t_{1}$ and $t_{2}$,
etc. The trace of each quantum trajectory gives its occurrence probability
(density), e.g., the jumpless trajectory occurs with a probability
$P_{\oslash}\equiv\operatorname*{Tr}\tilde{\rho}_{\oslash}$, the trajectory
$\tilde{\rho}_{t_{1}}$ occurs with a probability density $p_{t_{1}}%
\equiv\operatorname*{Tr}\tilde{\rho}_{t_{1}}$, the trajectory $\tilde{\rho
}_{t_{1}t_{2}}$ occurs with a probability density $p_{t_{1}t_{2}}%
\equiv\operatorname*{Tr}\tilde{\rho}_{t_{1}t_{2}}$, etc. The normalization
$\operatorname*{Tr}\bar{\rho}=1$ leads to the normalization of all the
probabilities:
\[
P_{\oslash}+\int_{0}^{T}p_{t_{1}}dt_{1}+\int_{0}^{T}dt_{2}\int_{0}^{t_{2}%
}dt_{1}\ p_{t_{1}t_{2}}+\cdots=1.
\]
These analytical expressions for the quantum trajectories in terms of the
quantum jump superoperator $\mathcal{J}$ and jumpless evolution superoperator
$\mathcal{L}_{0}$ correspond to an exact integration of the stochastic master
equation \cite{AlbarelliNJP2017,AlbarelliQ2018}.

In MQT, we not only track the quantum trajectory, but also record the timing
of every observed quantum jump during the evolution [see Fig. \ref{G_SCHEME}%
(b)]. At the end of the evolution, we make an optimal measurement on the
quantum probe to transfer all the QFI in the quantum probe into the
measurement outcome. After repeating this measurement cycle for $\nu\gg1$
times, we can use all the observed timings and the $\nu$ measurement outcomes
to construct an optimal unbiased estimator to $\theta$ (see Appendix A). The
fundamental precision is given by Eq. (\ref{DW_MQT}), where $\mathbb{F}%
=F+\mathcal{\bar{F}}$ [cf. Eq. (\ref{FMQT})] is the sum of the CFI [cf. Eq.
(\ref{CFI_TRAJ})]%
\begin{equation}
F=\frac{\left(  \partial_{\theta}P_{\oslash}\right)  ^{2}}{P_{\oslash}}%
+\int_{0}^{T}\frac{(\partial_{\theta}{p}_{t_{1}})^{2}}{p_{t_{1}}}dt_{1}%
+\cdots\label{CFI}%
\end{equation}
contained in all the timings of the quantum jumps and the trajectory-averaged
QFI:\
\begin{equation}
\mathcal{\bar{F}}\equiv P_{\oslash}\mathcal{F}[\rho_{\oslash}]+\int_{0}%
^{T}p_{t_{1}}\mathcal{F}[\rho_{t_{1}}]dt_{1}+\cdots, \label{FBAR}%
\end{equation}
where $\rho_{\oslash}\equiv\tilde{\rho}_{\oslash}/P_{\oslash}$, $\rho_{t_{1}%
}\equiv\tilde{\rho}_{t_{1}}/p_{t_{1}}$, etc. are normalized quantum trajectories.

For the estimation of Hamiltonian parameters, the jumpless trajectory
$\tilde{\rho}_{\oslash}$ is less (usually not) influenced by the decoherence,
so its QFI is much higher than other trajectories and the non-selective state
$\bar{\rho}$. When $[\mathcal{J},\mathcal{L}_{0}]=0$, we have $\tilde{\rho
}_{t_{1}\cdots t_{n}}\equiv\mathcal{J}^{n}\tilde{\rho}_{\oslash}$ and
$\bar{\rho}=e^{\mathcal{J}T}\tilde{\rho}_{\oslash}$, i.e., all the quantum
jumps can be deferred after the jumpless evolution. If $\mathcal{J}$ further
preserves the QFI, then all the quantum trajectories contain the same QFI as
the jumpless trajectory, so $\mathbb{F}\geq\mathcal{F}[\rho_{\oslash}]$ and
$\delta\theta_{\mathrm{MQT}}\leq1/\sqrt{\nu\mathcal{F}[\rho_{\oslash}]}$.

The capability of MQT to resolve different quantum trajectories motivates a
probabilistic protocol by post-selection of quantum trajectories. In this
protocol, after resolving the quantum trajectories, we only perform optimal
measurements on high-QFI trajectories and then construct an optimal unbiased
estimator based on the outcomes of these measurements, while discarding all
zero-QFI and even low-QFI trajectories. On one hand, this treatment reduces
the workload of performing a large number of optimal measurements and
constructing optimal unbiased estimators from a large number of measurement
outcomes (see Appendix A). On the other hand, discarding any trajectory with
nonzero QFI will degrade the fundamental estimation precision. Therefore, one
can balance between the estimation precision and the cost of measurement and
data post-processing to optimize the whole parameter estimation process. A
similar situation has been encountered in other probabilistic metrology
protocols such as weak-value amplification
\cite{KofmanPhysRep2012,DresselRMP2014}, where post-selection gains technical
advantages in data processing \cite{JordanPRX2014} at the cost of degrading
the fundamental estimation precision \cite{FerriePRL2014,CombesPRA2014}.

\subsection{Monitoring quantum trajectories via noiseless ancillas: connection
to QEC}

\label{SEC_QEC}

In most cases, MQT requires direct measurement of the noisy environment -- an
experimentally challenging task for realistic noise processes. Fortunately,
when the Hamiltonian evolution and the decoherence channel satisfy certain
conditions, it is possible to achieve MQT by quantum error encoding and error
detection assisted by noiseless ancillas
\cite{DuerPRL2014,KesslerPRL2014,ArradPRL2014,ZhouNC2018}, i.e., we can use
QEC to resolve the quantum trajectories, but do not apply any corrective
operations. Such QEC-based MQT can be regarded as a QEC protocol without
corrective operations, so it is less powerful than the full QEC-based
metrology
\cite{DuerPRL2014,KesslerPRL2014,ArradPRL2014,LuNC2015,HerreraMartiPRL2015,UndenPRL2016,BergmannPRA2016,MatsuzakiPRA2017,DemkowiczPRX2017,ZhouNC2018}%
. Nevertheless, MQT itself provides an interesting insight into how the
information leaks into the distinct quantum trajectories and how they are
recovered. Moreover, for very special cases where corrective operations are
not necessary (see Appendix B for an example), MQT becomes advantageous as it
avoids faulty corrective operations \cite{KesslerPRL2014}.

We begin with a simple example: monitoring the spin-flip channel of a spin-1/2
in the Lindblad form: $d\bar{\rho}(t)/dt=\gamma\mathcal{D}[S_{x}]\bar{\rho
}(t)$. Here the quantum jump operator $S_{x}$ induce random jumps between
$|\uparrow\rangle$ and $|\downarrow\rangle$. These random jumps can be
monitored via the QEC protocol \cite{KesslerPRL2014}, which adds an ancilla
that is not affected by the noise. Physically, this can be realized in a
nitrogen-vacancy center \cite{RondinRPP2014,SchirhaglARPC2014}, where the
electron spin serves as the quantum probe spin-1/2 and the $^{15}$N nuclear
spin serves as the ancilla. We use $|\uparrow\rangle$, $|\downarrow\rangle$
for the spin-1/2, $|0\rangle,|1\rangle$ for the ancilla, $\sigma_{z}$ for the
Pauli matrix on the spin-1/2, and $\sigma_{z}^{(a)}$ for the Pauli matrix on
the ancilla:\ $\sigma_{z}^{(a)}|0\rangle=|0\rangle$ and $\sigma_{z}%
^{(a)}|1\rangle=-|1\rangle$. The syndrome operator is $\Sigma\equiv\sigma
_{z}\sigma_{z}^{(a)}$. The code subspace spanned by $|\uparrow\rangle
|0\rangle$ and $|\downarrow\rangle|1\rangle$ is the eigensubspace of the
syndrome operator with eigenvalue $+1$, so we denote the code subspace by
$\Sigma_{+}$. The occurrence of a spin flip maps the code subspace $\Sigma
_{+}$ onto an orthogonal error subspace as spanned by $|\downarrow
\rangle|0\rangle$ and $|\uparrow\rangle|1\rangle$. This error subspace is an
eigensubspace of the syndrome operator $\Sigma$ with eigenvalue $-1$, so we
denote it by $\Sigma_{-}$. The occurrence of another spin flip maps
$\Sigma_{-}$ back to $\Sigma_{+}$. Therefore, frequently measuring the
syndrome operator $\Sigma$ allows us to monitor the spin flip in real time.
For Hamiltonian parameter estimation, if the Hamiltonian commutes with
$\Sigma$ and hence leaves $\Sigma_{\pm}$ invariant, then monitoring the spin
flip does not affect the coherent Hamiltonian evolution. As an example, we
consider $H=\omega S_{z}$ with $\omega$ the unknown parameter to be estimated.
The total non-selective evolution is $d\bar{\rho}(t)/dt=-i[H,\bar{\rho
}(t)]+\gamma\mathcal{D}[S_{x}]\bar{\rho}(t)$. Starting from an initial state
$|\psi_{0}\rangle=a|\uparrow\rangle|0\rangle+b|\downarrow\rangle|1\rangle
\in\Sigma_{+}$, in the absence of quantum jumps, the Hamiltonian evolution
keeps the state inside $\Sigma_{+}$: $|\psi(t)\rangle=a|\uparrow
\rangle|0\rangle+e^{i\omega t}b|\downarrow\rangle|1\rangle$. The occurrence of
a quantum jump at $t_{0}$ maps $|\psi(t_{0})\rangle$ to $|\tilde{\psi}%
(t_{0})\rangle=a|\downarrow\rangle|0\rangle+e^{i\omega t_{0}}b|\uparrow
\rangle|1\rangle\in\Sigma_{-}$, which can be detected as a sign switch of
$\Sigma$. The subsequent Hamiltonian evolution keeps the state inside
$\Sigma_{-}$: $|\tilde{\psi}(t)\rangle=a|\downarrow\rangle|0\rangle
+e^{i\omega(t_{0}-t)}b|\uparrow\rangle|1\rangle$. The occurrence of another
quantum jump at $t_{1}$ maps $|\tilde{\psi}(t_{1})\rangle$ back to the code
subspace: $|\psi(t_{1})\rangle=a|\uparrow\rangle|0\rangle+e^{i\omega
(t_{0}-t_{1})}b|\downarrow\rangle|1\rangle\in\Sigma_{+}$, which can be
detected as another sign switch of $\Sigma$. Here, in contrast to the full QEC
protocols
\cite{DuerPRL2014,KesslerPRL2014,ArradPRL2014,LuNC2015,HerreraMartiPRL2015,UndenPRL2016,BergmannPRA2016}%
, we only monitor the quantum trajectory without applying any corrective operations.

\begin{figure}[ptb]
\includegraphics[width=0.8\columnwidth]{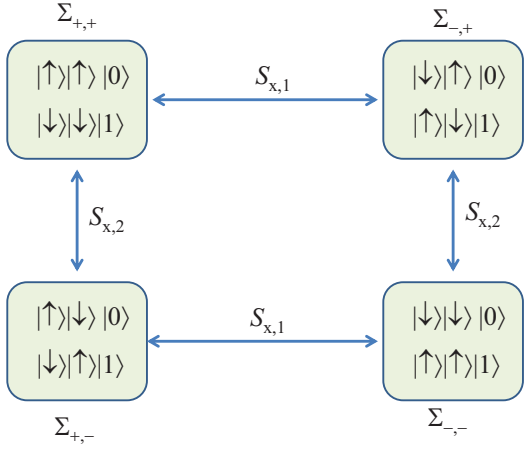}\caption{QEC-based MQT for two
spin-1/2's with the assistance of a noiseless ancilla. The four distinct
eigensubspaces of the two-component syndrome operator $\mathbf{\Sigma}%
=(\sigma_{z}^{(1)}\sigma_{z}^{(a)},\sigma_{z}^{(2)}\sigma_{z}^{(a)})$ defines
the code subspace $\Sigma_{+,+}$ and three error subspaces $\Sigma
_{-,+},\Sigma_{+,-},\Sigma_{-,-}$. Different subspaces are connected by the
flip of individual spins.}%
\label{G_QEC}%
\end{figure}

The key ingredients of this QEC-based MQT are: (i) The code subspace
$\Sigma_{+}$ and the error subspace $\Sigma_{-}$ are eigensubspaces of the
syndrome operator $\Sigma$ with distinct eigenvalues. (ii) The quantum jump
operator anti-commutes with the syndrome operator, so it maps an eigenstate of
the syndrome operator to another eigenstate with an opposite eigenvalue, i.e.,
it induces transition between $\Sigma_{+}$ and $\Sigma_{-}$ and hence can be
detected by measuring the syndrome operator \cite{NielsenBook2000}. If, in
addition, the Hamiltonian commutes with the syndrome operator, so that the
Hamiltonian evolution leaves $\Sigma_{+}$ and $\Sigma_{-}$ invariant, then the
detection of the quantum jump does not affect the Hamiltonian evolution,
similar to the full QEC-based metrology \cite{KesslerPRL2014,UndenPRL2016}.
For example, a single noiseless ancilla allows us to monitor the spin flip of
an arbitrary number $N$ of spin-1/2's by using the $N$-component syndrome
operator $\mathbf{\Sigma}\equiv(\sigma_{z}^{(1)}\sigma_{z}^{(a)},\cdots
,\sigma_{z}^{(N)}\sigma_{z}^{(a)})$, where $\sigma_{z}^{(i)}$ is the Pauli
operator for the $i$th spin-1/2 and the flip of the $i$th spin-1/2 is detected
as the sign switch of the $i$th component $\sigma_{z}^{(i)}\sigma_{z}^{(a)}$
(see Fig. \ref{G_QEC} for an example for $N=2$). When the Hamiltonian leaves
each eigensubspace invariant, e.g., $H=\omega(S_{z}^{(1)}+\cdots+S_{z}^{(N)}%
)$, MQT does not affect the Hamiltonian evolution.

The requirement that MQT leave the Hamiltonian evolution intact can be
satisfied only when the Hamiltonian is completely \textquotedblleft
transversal\textquotedblright\ to the quantum jump operator
\cite{DegenRMP2017} (e.g., $\omega S_{z}$ is transversal to $S_{x}$ in our
example). Generally, the Hamiltonian is the sum of a \textquotedblleft
transversal\textquotedblright\ component $H_{\perp}$ (i.e., the component
outside the Lindblad span of the decoherence channel
\cite{DemkowiczPRX2017,ZhouNC2018}) and a \textquotedblleft
parallel\textquotedblright\ component $H_{\parallel}$ (i.e., the component
inside the Lindblad span of the decoherence channel
\cite{DemkowiczPRX2017,ZhouNC2018}): the former keeps each eigensubspace of
the syndrome operator invariant, while the latter (as well as the quantum
jump) can induce transitions between different eigensubspaces. However,
frequent measurement on the syndrome operator leads to quantum Zeno effect
that effectively suppresses $H_{\parallel}$, so that only $H_{\perp}$ survives
\cite{ZhouNC2018}. In this case, MQT changes the intrinsic Hamiltonian
evolution and hence cannot reveal the intrinsic evolution of the quantum trajectories.

The discussions above suggest that for Hamiltonian parameter estimation, there
is an interesting connection between QEC-based MQT and the full QEC-based
metrology \cite{DemkowiczPRX2017,ZhouNC2018}: (i) When $H_{\perp}=0$, the
QEC-based MQT will completely freeze the Hamiltonian evolution, so the full
QEC protocol is not applicable to improve the estimation precision. (ii) When
$H_{\perp}\neq0$, the QEC-based MQT will suppress $H_{\parallel}$ but leave
$H_{\perp}$ intact, so the full QEC protocol can recover the Heisenberg
scaling of the estimation precision in the noiseless case
\cite{DemkowiczPRX2017,ZhouNC2018}. In particular, when $H_{\parallel}=0$, the
QEC-based MQT leaves the Hamiltonian evolution intact, so the full QEC
protocol can fully recover the estimation precision in the noiseless case
\cite{DemkowiczPRX2017,ZhouNC2018}.

\section{Application to spin-1/2}

We consider a spin-1/2 $\mathbf{S}$ undergoing the non-selective evolution
$\dot{\rho}(t)=\mathcal{L}\rho(t)$ starting from a general initial state
\[
\rho_{0}=%
\begin{bmatrix}
\rho_{\uparrow\uparrow} & \rho_{\uparrow\downarrow}\\
\rho_{\uparrow\downarrow}^{\ast} & \rho_{\downarrow\downarrow}%
\end{bmatrix}
\]
with $\rho_{\downarrow\downarrow}\equiv1-\rho_{\uparrow\uparrow}$, where
$\mathcal{L}\rho\equiv-i[\omega S_{z},{\rho}]+\gamma\mathcal{D}[c]\rho$,
$S_{z}$ is the $z$ component of the spin, $\omega$ is the level splitting, $c$
represents an arbitrary quantum jump operator, and $\gamma$ is the decoherence
rate. Monitoring the quantum jump $\mathcal{J}\rho\equiv\gamma c\rho c^{\dag}$
amounts to decomposing $\mathcal{L}$ into the quantum jump $\mathcal{J}$ and
the jumpless evolution $e^{\mathcal{L}_{0}t}\rho=e^{-i\tilde{H}t}\rho
e^{i\tilde{H}^{\dagger}t}$, where
\[
\tilde{H}\equiv\omega S_{z}-i\frac{\gamma}{2}c^{\dag}c
\]
is an effective non-Hermitian Hamiltonian that governs the jumpless trajectory
$\tilde{\rho}_{\oslash}=e^{\mathcal{L}_{0}T}\rho_{0}$. We consider the
estimation of the level splitting $\omega$ or the decoherence rate $\gamma$
(at $\omega=0$) under three decoherence channels: spin relaxation
$c=S_{-}=|\downarrow\rangle\langle\uparrow|$, spin flip $c=S_{x}$, and spin
dephasing $c=S_{z}$. We use a subscript $\omega$ ($\gamma$) to denote the
information about $\omega$ ($\gamma$), e.g., the CFI about $\omega$ ($\gamma
$)\ is $F_{\omega}$ ($F_{\gamma}$), and the total information about $\omega$
($\gamma$) from MQT is $\mathbb{F}_{\omega}$ ($\mathbb{F}_{\gamma}$).

For the spin-relaxation channel, QEC-based MQT is not possible, so MQT
requires direct measurement of the environment, which is experimentally
challenging for many realistic environments. For the special environment --
single-mode cavity, the spin relaxation due to the cavity field is always
accompanied by the emission of a cavity photon, so it can be monitored by
using a photon detector to measure the cavity output \cite{ZhengPRA2015}. For
the spin flip channel, the Hamiltonian $H=\omega S_{z}$ is completely
\textquotedblleft transversal\textquotedblright\ to the quantum jump $S_{x}$,
so QEC-based MQT is applicable and the syndrome operator is $\Sigma
\equiv\sigma_{z}\sigma_{z}^{(a)}$ (see Sec. \ref{SEC_QEC}). For the spin
dephasing channel, the Hamiltonian $H=\omega S_{z}$ completely lies in the
Lindblad span \cite{DemkowiczPRX2017,ZhouNC2018} of the decoherence channel,
i.e., $H_{\perp}=0$, so QEC-based MQT will completely freeze the Hamiltonian
evolution. Therefore, only when $\omega=0$ can the QEC-based MQT be used to
estimate $\gamma$ and, in this case, the results for $\mathbb{F}_{\gamma}$ are
identical to the spin flip channel since $S_{z}$ and $S_{x}$ are connected by
a unitary $\pi/2$-rotation around the $y$ axis. In other cases, MQT requires
directly monitoring the environment, which may be experimentally challenging.
In Appendix B, we give an example for using QEC-based MQT to recover the
Heisenberg scaling of frequency estimation for \textit{multiple} qubits.

In a recent work, Albarelli \textit{et al}. \cite{AlbarelliNJP2017} considered
the fundamental estimation precision of frequency by directly monitoring the
radiation field from an ensemble of atoms undergoing Markovian collective
dephasing. During the first-round revision of this manuscript after
submission, we became aware of another work by Albarelli \textit{et al}.
\cite{AlbarelliQ2018}, which further include the Markovian spin-flip channel
and demonstrated the interesting possibility of recovering the Heisenberg
scaling of the estimation precision with respect to the number $N$ of atoms.
These works focus on frequency estimation and its scaling with respect to the
number $N$ of atoms and rely on either the Gaussian approximation in the limit
of a large number of atoms \cite{AlbarelliNJP2017} or numerical integration of
the stochastic master equation \cite{AlbarelliQ2018}. Here we focus on the
time scaling of the estimation precision for both the frequency $\omega$ and
the dissipation rate $\gamma$ of a single spin-1/2. Moreover, the
superoperator formalism in Sec. \ref{SEC_SUPER} allows us\ to obtain explicit
analytical expressions.

In the following, we first give the quantum trajectories and their information
content for each decoherence channel and then discuss the estimation precision
for $\omega$ and $\gamma$.

\subsection{Quantum trajectories and Fisher information}

For the spin relaxation channel, $\tilde{H}=\omega S_{z}-i(\gamma
/2)|\uparrow\rangle\langle\uparrow|$ and $\mathcal{J}^{2}=0$, so the spin can
undergo at most one quantum jump. The jumpless trajectory is
\begin{equation}
\tilde{\rho}_{\oslash}=%
\begin{bmatrix}
e^{-\gamma T}\rho_{\uparrow\uparrow} & e^{-\gamma T/2}e^{-i\omega T}%
\rho_{\uparrow\downarrow}\\
e^{-\gamma T/2}e^{i\omega T}\rho_{\uparrow\downarrow}^{\ast} & \rho
_{\downarrow\downarrow}%
\end{bmatrix}
. \label{R0_SN}%
\end{equation}
The trajectory with an exclusive quantum jump at $t_{1}$ is $\tilde{\rho
}_{t_{1}}=\gamma e^{-\gamma t_{1}}\rho_{\uparrow\uparrow}|\downarrow
\rangle\langle\downarrow|$. Other trajectories $\tilde{\rho}_{t_{1}\cdots
t_{n}}$ with $n\geq2$ quantum jumps are absent. The sum of all the quantum
trajectories gives the non-selective density matrix%
\begin{equation}
{\bar{\rho}}=%
\begin{bmatrix}
e^{-\gamma T}\rho_{\uparrow\uparrow} & e^{-\gamma T/2}e^{-i\omega T}%
\rho_{\uparrow\downarrow}\\
e^{-\gamma T/2}e^{i\omega T}\rho_{\uparrow\downarrow}^{\ast} & 1-\rho
_{\uparrow\uparrow}e^{-\gamma T}%
\end{bmatrix}
. \label{RNS_SN}%
\end{equation}

For the spin flip channel, $\tilde{H}=\omega S_{z}-i(\gamma/8)$ and
$\mathcal{J}^{2}=(\gamma/4)^{2}$, so an arbitrary number of quantum jumps is
possible. The jumpless trajectory is%
\begin{equation}
\tilde{\rho}_{\oslash}=e^{-\gamma T/4}%
\begin{bmatrix}
\rho_{\uparrow\uparrow} & e^{-i\omega T}\rho_{\uparrow\downarrow}\\
e^{i\omega T}\rho_{\uparrow\downarrow}^{\ast} & \rho_{\downarrow\downarrow}%
\end{bmatrix}
. \label{R0_SX}%
\end{equation}
The trajectory with $n$ quantum jumps at $t_{1},\cdots,t_{n}$ is
\begin{equation}
\tilde{\rho}_{t_{1}\cdots t_{n}}=\left(  {\frac{\gamma}{4}}\right)
^{n}e^{-{\gamma T/4}}%
\begin{bmatrix}
\rho_{\uparrow\uparrow} & \rho_{\uparrow\downarrow}e^{-i\omega\int_{0}%
^{T}s(t)dt}\\
\rho_{\uparrow\downarrow}^{\ast}e^{i\omega\int_{0}^{T}s(t)dt} & \rho
_{\downarrow\downarrow}%
\end{bmatrix}
\label{RN_SX}%
\end{equation}
for even $n$ and $\sigma_{x}\tilde{\rho}_{t_{1}\cdots t_{n}}\sigma_{x}$ for
odd $n$, where $\sigma_{x}$ is the Pauli matrix and $s(t)$ starts from $+1$ at
$t=0$ and reverses its sign at $t_{1},\cdots,t_{n}$. The sum of all the
trajectories gives the non-selective final state%
\begin{equation}
\bar{\rho}=%
\begin{bmatrix}
\frac{1}{2}+{\frac{e^{-\gamma T/2}}{2}(}\rho_{\uparrow\uparrow}-\rho
_{\downarrow\downarrow}) & ({f\rho_{\uparrow\downarrow}+g\rho_{\uparrow
\downarrow}^{\ast})}e^{-\gamma T/4}\\
({f^{\ast}\rho_{\uparrow\downarrow}^{\ast}+g\rho_{\uparrow\downarrow}%
)}e^{-\gamma T/4} & \frac{1}{2}-{\frac{e^{-\gamma T/2}}{2}(}\rho
_{\uparrow\uparrow}-\rho_{\downarrow\downarrow})
\end{bmatrix}
, \label{RNS_SX}%
\end{equation}
where $f=\cos(wT)-i(\omega/w)\sin(wT)$ and $g=\gamma\sin{(wT)}/(4w)$ with
$w\equiv\sqrt{\omega^{2}-(\gamma/4)^{2}}.$

For the spin dephasing channel, $\tilde{H}=\omega S_{z}-i(\gamma/8)$ and
$\mathcal{J}^{2}=(\gamma/4)^{2}$, so an arbitrary number of quantum jumps is
possible. The jumpless trajectory is the same as the spin flip channel. The
trajectory with $n$ quantum jumps at $t_{1},\cdots,t_{n}$ is%
\begin{equation}
\tilde{\rho}_{t_{1}\cdots t_{n}}=\left(  {\frac{\gamma}{4}}\right)
^{n}e^{-{\gamma T/4}}%
\begin{bmatrix}
\rho_{\uparrow\uparrow} & (-1)^{n}\rho_{\uparrow\downarrow}e^{-i\omega T}\\
(-1)^{n}\rho_{\uparrow\downarrow}^{\ast}e^{i\omega T} & \rho_{\downarrow
\downarrow}%
\end{bmatrix}
, \label{RN_SZ}%
\end{equation}
which is independent of the timings of the $n$ quantum jumps. Summing all the
trajectories gives the non-selective final state%
\begin{equation}
\bar{\rho}=%
\begin{bmatrix}
\rho_{\uparrow\uparrow} & e^{-\gamma T/2}e^{-i\omega T}\rho_{\uparrow
\downarrow}\\
e^{-\gamma T/2}e^{i\omega T}\rho_{\uparrow\downarrow}^{\ast} & \rho
_{\downarrow\downarrow}%
\end{bmatrix}
. \label{RNS_SZ}%
\end{equation}

In Table \ref{QFI_ALL}, we list the information for estimating $\omega$ and
for estimating $\gamma$ at $\omega=0$ for each decoherence channel. In the MQT
approach, after $\nu$ repeated measurement cycles, the estimation precision
\begin{equation}
\delta\theta=\frac{1}{\sqrt{\nu\mathbb{F}_{\theta}}} \label{DT_MQT}%
\end{equation}
is determined by the total information $\mathbb{F}_{\theta}$ ($\theta=\omega$
or $\gamma$) from each measurement cycle. In the conventional approach, after
$\nu$ repeated measurement cycles, the estimation precision
\begin{equation}
\delta\bar{\theta}=\frac{1}{\sqrt{\nu\mathcal{F}_{\theta}[\bar{\rho}]}}
\label{DT_C}%
\end{equation}
is determined by the QFI $\mathcal{F}_{\theta}[\bar{\rho}]$ ($\theta=\omega$
or $\gamma$) of the non-selective final state $\bar{\rho}$ in each measurement
cycle. According to Eq.\ (\ref{INEQ}), we always have $\mathbb{F}_{\theta}%
\geq\mathcal{F}_{\theta}[\bar{\rho}]$, so MQT always improve the estimation
precision, but the degree of improvement depend on the parameter to be
estimated and the decoherence channel.

\begin{widetext}
\renewcommand\arraystretch{2.5}
\begin{table*}[tbp] \centering
\begin{tabular}
[c]{cccc}\hline\hline
Information about $\omega$ & Spin relaxation channel ($c=S_{-}$)\  & Spin flip
channel ($c=S_{x}$) & Spin dephasing\ channel ($c=S_{z}$)\ \\\hline
$\mathcal{F}_{\omega}[\rho_{\oslash}]$ & ${\dfrac{4T^{2}|\rho_{\uparrow\downarrow
}|^{2}e^{-\gamma T}}{(\rho_{\uparrow\uparrow}e^{-\gamma T}+\rho_{\downarrow
\downarrow})^{2}}}$ & $4T^{2}|\rho_{\uparrow\downarrow}|^{2}$ & $4T^{2}%
|\rho_{\uparrow\downarrow}|^{2}$\\
$\mathcal{F}_{\omega}[\rho_{t_{1}\cdots t_{n}}]$ \  & $0$ & $4\left[  \int
_{0}^{T}s(t)dt\right]  ^{2}|\rho_{\uparrow\downarrow}|^{2}$ & $4T^{2}%
|\rho_{\uparrow\downarrow}|^{2}$\\
$F_{\omega}$ & 0 & 0 & 0\\
$\mathbb{F}_{\omega}=\mathcal{\bar{F}}_{\omega}$ & ${\dfrac{4T^{2}%
|\rho_{\uparrow\downarrow}|^{2}e^{-\gamma T}}{\rho_{\uparrow\uparrow
}e^{-\gamma T}+\rho_{\downarrow\downarrow}}}$ & ${\dfrac{32|\rho
_{\uparrow\downarrow}|^{2}}{\gamma^{2}}}\left(  {\dfrac{\gamma T}{2}%
}-(1-e^{-{\gamma T/2}})\right)  $ & $4T^{2}|\rho_{\uparrow\downarrow}|^{2}$\\
$\mathcal{F}_{\omega}[\bar{\rho}]$ & $4T^{2}|\rho_{\uparrow\downarrow
}|^{2}e^{-\gamma T}$ & $4T^{2}|\rho_{\uparrow\downarrow}|^{2}e^{-\gamma T/2}$
& $4T^{2}|\rho_{\uparrow\downarrow}|^{2}e^{-\gamma T}$\\\hline\hline
Information about $\gamma$ at $\omega=0$ & Spin relaxation channel ($c=S_{-}$)\  & Spin flip
channel ($c=S_{x}$) & Spin dephasing\ channel ($c=S_{z}$)\ \\\hline
$\mathcal{F}_{\gamma}[\rho_{\oslash}]$ & ${\dfrac{T^{2}\rho_{\uparrow\uparrow}%
\rho_{\downarrow\downarrow}e^{-\gamma T}}{(\rho_{\uparrow\uparrow}e^{-\gamma
T}+\rho_{\downarrow\downarrow})^{2}}}$ & $0$ & $0$\\
$\mathcal{F}_{\gamma}[\rho_{t_{1}\cdots t_{n}}]$ \  & $0$ & $0$ & $0$\\
$F_{\gamma}$ & ${\dfrac{\rho_{\uparrow\uparrow}(1-e^{-\gamma T})}{\gamma^{2}}%
}-{\dfrac{T^{2}\rho_{\uparrow\uparrow}\rho_{\downarrow\downarrow}e^{-\gamma
T}}{\rho_{\uparrow\uparrow}e^{-\gamma T}+\rho_{\downarrow\downarrow}}}$ &
${\dfrac{T}{4\gamma}}$ & ${\dfrac{T}{4\gamma}}$\\
$\mathbb{F}_{\gamma}$ & ${\dfrac{\rho_{\uparrow\uparrow}}{\gamma^{2}}%
}(1-e^{-\gamma T})$ & ${\dfrac{T}{4\gamma}}$ & ${\dfrac{T}{4\gamma}}$\\
$\mathcal{F}_{\gamma}[\bar{\rho}]$ & $\dfrac{T^{2}\rho_{\uparrow
\uparrow}\left[  \rho_{\uparrow\uparrow}-|\rho_{\uparrow\downarrow}|^{2}%
(\rho_{\uparrow\uparrow}e^{-\gamma T}+1)\right]  e^{-\gamma T}}{\rho
_{\uparrow\uparrow}(1-\rho_{\uparrow\uparrow}e^{-\gamma T})-|\rho
_{\uparrow\downarrow}|^{2}}$ & $\dfrac{T^{2}\left\vert \rho_{+-}\right\vert
^{2}e^{-\gamma T}\rho_{++}\rho_{--}}{\rho_{++}\rho_{--}-\left\vert \rho
_{+-}\right\vert ^{2}e^{-\gamma T}}$ & ${\dfrac{T^{2}|\rho_{\uparrow
\downarrow}|^{2}e^{-\gamma T}\rho_{\uparrow\uparrow}\rho_{\downarrow
\downarrow}}{\rho_{\uparrow\uparrow}\rho_{\downarrow\downarrow}-|\rho
_{\uparrow\downarrow}|^{2}e^{-\gamma T}}}$\\\hline\hline
\end{tabular}
\caption{Information about the qubit frequency $\omega$ or the decoherence rate $\gamma$ (at $\omega=0$) under different decoherence channels, where $s(t)$ starts from +1 at $t=0$ and switches its sign at every quantum jump $(t_1,\cdots, t_n)$, $\rho_{++},\rho_{--},\rho_{+-}$ are matrix elements of the initial state $\rho_0$ in the basis $|\pm\rangle\equiv\left(  |\uparrow\rangle\pm|\downarrow\rangle\right)  /\sqrt{2}$. The expression for $\mathcal{F}_{\omega}[\bar{\rho}]$ under  the spin flip channel is valid up to leading order of $\gamma/\omega$.}\label{QFI_ALL}%
\end{table*}%
\renewcommand\arraystretch{1.0}
\end{widetext}

From Table \ref{QFI_ALL}, we see that estimating the Hamiltonian parameter
$\omega$ is very different from estimating the decoherence parameter $\gamma$
(at $\omega=0$):

(1) For all the decoherence channels, the CFI about $\omega$ ($\gamma$) is
zero (nonzero), i.e., the timings of the quantum jumps contain information
about $\gamma$, but no information about $\omega$. Physically, $\gamma$
characterizes the rate of the quantum jump, so the occurrence probabilities of
all the quantum trajectories depend on $\gamma$, but are independent of
$\omega$, leading to nonzero $F_{\gamma}$ but vanishing $F_{\omega}$ according
to Eq. (\ref{CFI}). Therefore, for estimating $\omega$, we need not record the
timings of the quantum jumps.

(2) For the spin flip and spin dephasing channels, the QFI about $\omega$
($\gamma$) is nonzero (zero) for all the trajectories. Physically, the
dependence of the \textit{normalized} quantum trajectories on $\omega$ and
$\gamma$ originate from the jumpless evolution $\mathcal{L}_{0}$ or
equivalently $\tilde{H}=\omega S_{z}-i(\gamma/8)$, which imprints the $\omega$
dependence but no $\gamma$ dependence onto the normalized quantum
trajectories. Therefore, for estimating $\gamma$, we need only record the
timings of the quantum jumps, while measurements over the final state are not necessary.

(3)\ For the spin relaxation channel, the trajectory with quantum jumps has
vanishing QFI about $\omega$ and $\gamma$, because a single quantum jump
projects an arbitrary state into $|\downarrow\rangle$ and hence eliminate all
the information. Therefore, once a quantum jump is detected, we can
immediately stop the current measurement cycle and start the next measurement
cycle to reduce the total time cost.

(4) The information about $\omega$ are all proportional to $|\rho
_{\uparrow\downarrow}|^{2}$, i.e., the projection of the initial spin in the
$xy$ plane, because $\omega$ is imprinted onto the quantum probe through the
Larmor precession around the $z$ axis. The positive-definiteness of the
density matrix dictates $|\rho_{\uparrow\downarrow}|^{2}\leq\rho
_{\uparrow\uparrow}\rho_{\downarrow\downarrow}$, so in the following we set
$|\rho_{\uparrow\downarrow}|^{2}=\rho_{\uparrow\uparrow}\rho_{\downarrow
\downarrow}$ to optimize the estimation precision for $\omega$.

\subsection{Estimation of $\omega$}

For the spin relaxation channel, the jumpless trajectory [Eq. (\ref{R0_SN})]
contains the QFI
\[
\mathcal{F}_{\omega}[\rho_{\oslash}]=4T^{2}{\frac{\rho_{\uparrow\uparrow
}e^{-\gamma T}\rho_{\downarrow\downarrow}}{(\rho_{\uparrow\uparrow}e^{-\gamma
T}+\rho_{\downarrow\downarrow})^{2}}}%
\]
and occurs with a probability $P_{\oslash}(T)=\rho_{\uparrow\uparrow
}e^{-\gamma T}+\rho_{\downarrow\downarrow}$, while the trajectories with a
quantum jump contains no QFI. Given an arbitrary evolution time $T$, preparing
the initial state $\rho_{\downarrow\downarrow}=1-\rho_{\uparrow\uparrow
}=1/(e^{\gamma T}+1)$ makes the QFI of the jumpless trajectory attains its
maximum $T^{2}$. This motivates a parameter estimation protocol based on the
probabilistic preparation of the jumpless trajectory (which contains
\textit{all} the information about $\omega$): $\nu$ successful preparation of
the jumpless trajectory herald the estimation precision $\delta\omega
=1/(\sqrt{\nu}T)$, which attains the Heisenberg scaling. The drawback is that
the success probability $P_{\oslash}(T)=2/(e^{\gamma T}+1)$ for each
preparation decreases with increasing $T$. As discussed in Sec.
\ref{SEC_SUPER}, such probabilistic protocol reduces the workload of data
processing at the cost of reducing the fundamental estimation precision.

\begin{figure}[ptb]
\includegraphics[width=\columnwidth]{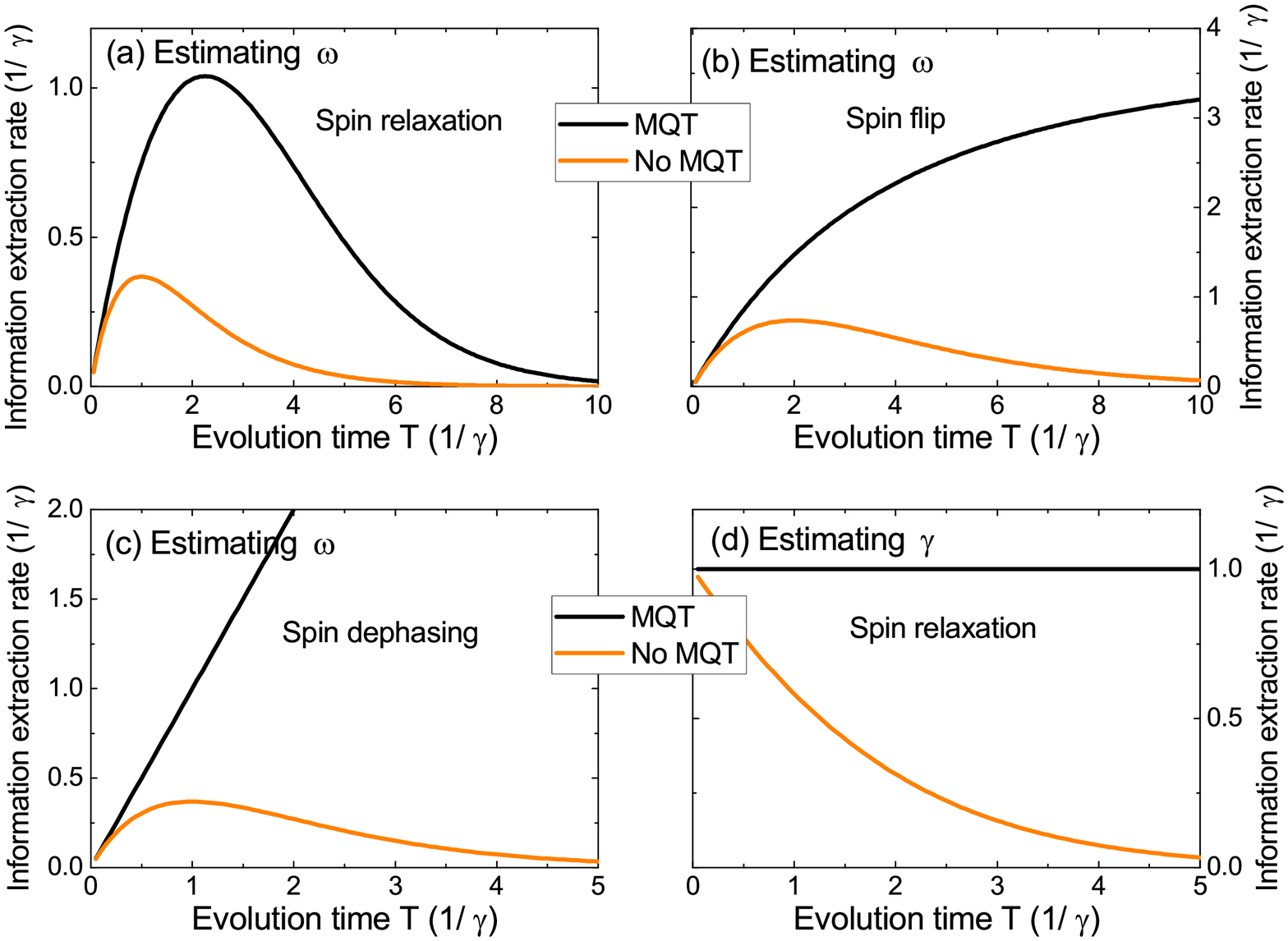}\caption{Optimal
information extraction rates (obtained by optimizing the initial states) for
estimating $\omega$ (a)-(c) or $\gamma$ (d) for different decoherence
channels: black lines for MQT and orange lines for the conventional method. }%
\label{G_RATE}%
\end{figure}

For deterministic parameter estimation, we should take into account the
occurrence probabilities of the quantum trajectories. The estimation precision
of MQT is determined by
\[
\mathbb{F}_{\omega}={\frac{\mathcal{F}_{\omega}[\bar{\rho}]}{\rho
_{\uparrow\uparrow}e^{-\gamma T}+\rho_{\downarrow\downarrow}},}%
\]
where $\mathcal{F}_{\omega}[\bar{\rho}]=4T^{2}\rho_{\uparrow\uparrow
}e^{-\gamma T}\rho_{\downarrow\downarrow}$ is the QFI of the conventional
approach. For small $\rho_{\downarrow\downarrow}$, the enhancement of
$\mathbb{F}_{\omega}$ relative to $\mathcal{F}_{\omega}[\bar{\rho}]$ could be
very large when $\gamma T\gg1$. In addition, MQT also shortens the total time
cost:\ once a quantum jump is detected, we should immediately stop the current
measurement cycle (because the trajectory with a quantum jump has no QFI) and
start the next measurement cycle. In this case, the average time cost for each
measurement cycle is%
\begin{equation}
T_{\mathrm{ave}}=P_{\oslash}T+\int_{0}^{T}t_{1}\operatorname*{Tr}\tilde{\rho
}_{t_{1}}dt_{1}\newline=\frac{\rho_{\uparrow\uparrow}}{\gamma}(1-e^{-\gamma
T})+\rho_{\downarrow\downarrow}T, \label{T_AVE}%
\end{equation}
which is shorter than $T$, especially when $\rho_{\downarrow\downarrow}$ is
small. Since the total time cost of $\nu$ repeated measurement cycles is $\nu
T_{\mathrm{ave}}$, the estimation precision per unit time, i.e., the
sensitivity $\delta\omega\sqrt{\nu T_{\mathrm{ave}}}=1/\sqrt{\mathbb{F}%
_{\omega}/T_{\mathrm{ave}}}$ [cf. Eq. (\ref{DT_MQT})], is determined by the
information extraction rate
\[
\frac{\mathbb{F}_{\omega}}{T_{\mathrm{ave}}}\leq\frac{4\gamma T^{2}}%
{(\sqrt{e^{\gamma T}-1}+\sqrt{\gamma T})^{2}}\overset{\gamma T\gg
1}{\longrightarrow}4\gamma T^{2}e^{-\gamma T},
\]
where the inequality is saturated at $\rho_{\downarrow\downarrow}%
/\rho_{\uparrow\uparrow}=\sqrt{(1-e^{-\gamma T})/(\gamma Te^{\gamma T})}$. For
comparison, the information extracting rate of the conventional method is
\[
\frac{\mathcal{F}_{\omega}[\bar{\rho}]}{T}\leq Te^{-\gamma T},
\]
where the inequality is saturated at $\rho_{\downarrow\downarrow}%
=\rho_{\uparrow\uparrow}=1/2$. Therefore, MQT enhances the long-time
information extraction rate by a factor $4\gamma T$ although both
$\mathbb{F}_{\omega}$ and $\mathcal{F}_{\omega}[\bar{\rho}]$ exhibit the same
exponential decay, as shown in Fig. \ref{G_RATE}(a).

For the spin flip channel, the jumpless trajectory [Eq. (\ref{R0_SX})]
contains the QFI
\[
\mathcal{F}_{\omega}[\rho_{\oslash}]=4T^{2}\rho_{\uparrow\uparrow}%
\rho_{\downarrow\downarrow}%
\]
that attains the Heisenberg scaling with respect to $T$, while the trajectory
with $n$ quantum jumps at $t_{1},\cdots,t_{n}$ [Eq. (\ref{RN_SX})] contains
the QFI
\[
\mathcal{F}_{\omega}[\rho_{t_{1}\cdots t_{n}}]=4\left(  \int_{0}%
^{T}s(t)dt\right)  ^{2}\rho_{\uparrow\uparrow}\rho_{\downarrow\downarrow},
\]
where $s(t)$ starts from $+1$ at $t=0$ and reverses its sign at $t_{1}%
,\cdots,t_{n}$. Physically, every quantum jump reverses the direction of the
phase accumulation [manifested as the sign reversal of $s(t)$] and $\int
_{0}^{T}s(t)dt$ is the \textquotedblleft net\textquotedblright\ phase
accumulation time. For example, we consider the trajectory with a single
quantum jump at $t_{1}$. Starting from an initial state $a|\uparrow
\rangle+b|\downarrow\rangle$, the quantum probe first evolves as
$a|\uparrow\rangle+e^{i\omega t_{1}}b|\downarrow\rangle$ and then undergoes a
spin flip into the state $a|\downarrow\rangle+e^{i\omega t_{1}}b|\uparrow
\rangle$, and then evolve into the final state $a|\downarrow\rangle
+e^{-i\omega(T-t_{1})}e^{i\omega t_{1}}b|\uparrow\rangle$, so its
\textquotedblleft net\textquotedblright\ phase accumulation time is
$t_{1}-(T-t_{1})$, which coincides with $\int_{0}^{T}s(t)dt$.

Here the Heisenberg scaling of $\mathcal{F}_{\omega}[\rho_{\oslash}]$ again
motivates a probabilistic parameter estimation protocol based on preparing the
jumpless trajectory: $\nu$ successful preparations with evolution time $T$ and
$\rho_{\uparrow\uparrow}=\rho_{\downarrow\downarrow}=1/2$ herald the
estimation precision $\delta\omega=1/(\sqrt{\nu}T)$, which attains the
Heisenberg scaling with respect to $T$. The drawback is that the success
probability $P_{\oslash}(T)=e^{-\gamma T/4}$ for each preparation decreases
with increasing $T$, so this probabilistic protocol reduces the workload of
data processing at the cost of reducing the fundamental estimation precision.
For deterministic parameter estimation, we average the QFIs of every
trajectory over their occurrence probabilities to obtain%
\[
\frac{\mathbb{F}_{\omega}}{T}\overset{\gamma T\gg1}{\longrightarrow}%
{\frac{{16}}{\gamma}\rho_{\uparrow\uparrow}\rho_{\downarrow\downarrow}.}%
\]
For comparison, the information extraction rate of the conventional method is
\[
\frac{\mathcal{F}_{\omega}[\bar{\rho}]}{T}=4T\rho_{\uparrow\uparrow}%
\rho_{\downarrow\downarrow}e^{-\gamma T/2}.
\]
Thus MQT avoids the long-time exponential decay of the sensitivity (i.e.,
precision per unit time), as shown in Fig. \ref{G_RATE}(b). Physically, the
jumpless trajectory has a negligible occurrence probability, so the long-time
linear scaling $\mathbb{F}_{\omega}=\mathcal{\bar{F}}_{\omega}\propto T$ comes
from the QFIs of other trajectories: the random quantum jumps leads to random
sign reversal of $s(t)$, so $[\int_{0}^{T}s(t)dt]^{2}$ and hence the QFI of
each trajectory increase linearly with $T$ on average, similar to a random
walk. MQT gives access to this trajectory-resolved QFI, as opposed to the
ensemble QFI $\mathcal{F}_{\omega}[\bar{\rho}]$ from the conventional method.

When we take into account the finite interval $\Delta$ between successive
syndrome measurements in QEC-based MQT for the spin-flip channel, we obtain%
\begin{equation}
\mathbb{F}_{\omega}=\frac{32\rho_{\uparrow\uparrow}\rho_{\downarrow\downarrow
}}{\gamma^{2}}e^{-\gamma T/4}e^{2\zeta}\left[  (1-\zeta e^{-\zeta}%
)^{N}+(1+\zeta e^{-\zeta})^{N}\left(  \frac{\gamma T}{2}e^{-\zeta}-1\right)
\right]  \label{FW_DELTA}%
\end{equation}
under the condition $1/\Delta\gg\omega\gg\gamma$ and\textit{ }$\gamma
T\ll1/(\gamma\Delta)$, where $\zeta\equiv\gamma\Delta/4$ and $N\equiv
T/\Delta$ is the total number of syndrome measurements during the interval
$T$. When the syndrome measurements are much faster than the decoherence
($\zeta\ll1$), $\mathbb{F}_{\omega}$ approaches the ideal results in Table
\ref{QFI_ALL}. When each syndrome measurement is imperfect, we can combine
several adjacent imperfect syndrome measurements into a composite measurement
to suppress the measurement error exponentially.

Finally, we turn to the spin dephasing channel. Interestingly, each quantum
jump merely flips the phase between $|\uparrow\rangle$ and $|\downarrow
\rangle$ without affecting the phase accumulation [Eq. (\ref{RN_SZ})], so
every quantum trajectory contain the same QFI
\begin{equation}
\mathbb{F}_{\omega}=\mathcal{F}_{\omega}[\rho_{\oslash}]=\mathcal{F}_{\omega
}[\rho_{t_{1}\cdots t_{n}}]=4T^{2}\rho_{\uparrow\uparrow}\rho_{\downarrow
\downarrow}, \label{FW_SZ}%
\end{equation}
in contrast to the conventional method:%
\[
\mathcal{F}_{\omega}[\bar{\rho}]=4T^{2}\rho_{\uparrow\uparrow}\rho
_{\downarrow\downarrow}e^{-\gamma T}.
\]
So MQT can restore the Heisenberg scaling of the estimation precision, as
shown in Fig. \ref{G_RATE}(c). Unfortunately, QEC-based MQT for the dephasing
channel of a single spin-1/2 is not possible, so direct measurement over the
environment is necessary. For \textit{multiple} qubits, the key to the
recovery of the Heisenberg scaling for estimating $\omega$ is $[\mathcal{J}%
,\mathcal{L}_{0}]=0$ and the preservation of the QFI\ under the quantum jump
$\mathcal{J}$ [see the discussions after Eq. (\ref{CFI})]. When this condition
is satisifed, QEC-based MQT can be used to recover the Heisenberg scaling (see
Appendix B for an example).

\subsection{Estimation of $\gamma$}

For the spin relaxation channel, the estimation precision of MQT is determined
by $\mathbb{F}_{\gamma}\approx\rho_{\uparrow\uparrow}/\gamma^{2}$ at $\gamma
T\gg1$. By contrast, the estimation precision of the conventional method is
determined by $\mathcal{F}_{\gamma}[\bar{\rho}]\approx{T^{2}\rho
_{\uparrow\uparrow}e^{-\gamma T}}$ at $\gamma T\gg1$. Thus MQT avoids the
exponential loss of the QFI in the conventional method. In addition, it also
shortens the total time cost from $T$ to $T_{\mathrm{ave}}$ [Eq.
(\ref{T_AVE})]. The information extraction rate, which determines the
sensitivity $\delta\gamma\sqrt{\nu T_{\mathrm{ave}}}=1/\sqrt{\mathbb{F}%
_{\gamma}/T_{\mathrm{ave}}}$, is
\begin{equation}
\frac{\mathbb{F}_{\gamma}}{T_{\mathrm{ave}}}\newline\leq\frac{1}{\gamma
},\label{RATE_SN}%
\end{equation}
where the inequality is saturated at $\rho_{\downarrow\downarrow}=0$. The
information extracting rate of the conventional method is
\begin{equation}
\frac{\mathcal{F}_{\gamma}[\bar{\rho}]}{T}\leq\frac{T}{e^{\gamma T}%
-1},\label{RATENS_SN}%
\end{equation}
where the inequality is saturated at $\rho_{\uparrow\downarrow}=\rho
_{\downarrow\downarrow}=0$. Therefore, MQT avoids the long-time exponential
loss of the sensitivity, as shown in Fig. \ref{G_RATE}(d).

For the spin flip channel, the information about $\gamma$ is entirely the CFI
in the timings of the quantum jumps: $\mathbb{F}_{\gamma}=F_{\gamma
}=T/(4\gamma)$. The information extraction rate from the MQT is
\begin{equation}
\frac{\mathbb{F}_{\gamma}}{T}=\frac{1}{4\gamma}. \label{RATEG_MQT}%
\end{equation}
The information extraction rate of the conventional method is
\begin{equation}
\frac{\mathcal{F}_{\gamma}[\bar{\rho}]}{T}\leq\frac{T}{4(e^{\gamma T}-1)},
\label{RATEG_C}%
\end{equation}
where $\rho_{++},\rho_{--}$ are the populations of $|\pm\rangle\equiv
(|\uparrow\rangle\pm|\downarrow\rangle)/\sqrt{2}$ in the initial state
$\rho_{0}$ and the inequality is saturated at $|\rho_{+-}|^{2}=\rho_{++}%
\rho_{--}$ and $\rho_{++}=\rho_{--}=1/2$. Equation (\ref{RATEG_C}) is also the
ultimate precision bound \cite{PirandolaPRL2017} for adaptive estimation of
the dephasing rate without MQT. Comparing Eq. (\ref{RATEG_MQT}) to Eq.
(\ref{RATEG_C}), we see that MQT avoids the long-time exponential loss of the
sensitivity. At $\omega=0$, including the finite interval $\Delta$ between
successive syndrome measurements in QEC-based MQT for the spin-flip channel
amounts to a multiplicative factor $4\zeta/(e^{4\zeta}-1)$ (with $\zeta
\equiv\gamma\Delta/4$ and $N\equiv T/\Delta$) to $\mathbb{F}_{\gamma}$. When
$\zeta\ll1$, $\mathbb{F}_{\gamma}$ approaches the ideal results in Table
\ref{QFI_ALL}. For the spin dephasing channel, we obtain exactly the same
results, with $|\uparrow\rangle$ ($|\downarrow\rangle$) playing the role of
$|+\rangle$ ($|-\rangle)$.

\section{Conclusion}

Quantum enhanced parameter estimation has widespread applications in many
fields. An important issue is to protect the estimation precision against the
noise-induced decoherence. For this purpose, we have developed a general
theoretical framework for improving the precision of parameter estimation by
monitoring the noise-induced quantum trajectories (MQT) of the quantum probe
and further establish its connection to the purification-based approach to
quantum parameter estimation \cite{FujiwaraJPA2008}. For Markovian
environment, we provide a superoperator approach to determining the
fundamental bounds on the estimation precision. This approach may provide
exact analytical expressions for some simple models. MQT can be achieved in
two ways: (i)\ Any quantum trajectories can be monitored by directly
monitoring the environment, which is experimentally challenging for realistic
noises. (ii) Certain quantum trajectories can also be monitored by frequently
measuring the quantum probe (without monitoring the environment) via
ancilla-assisted encoding and error detection, as used in quantum error
correction (QEC). This QEC-based MQT makes certain MQT feasible and further
establishes an interesting connection between MQT and the metrology protocols
based on full QEC
\cite{DuerPRL2014,KesslerPRL2014,ArradPRL2014,LuNC2015,HerreraMartiPRL2015,UndenPRL2016,BergmannPRA2016,MatsuzakiPRA2017}%
. We apply MQT to the estimation of the level splitting $\omega$ and the
decoherence rate $\gamma$ of a spin-1/2 under three decoherence channels: spin
relaxation, spin flip, and spin dephasing. We find that it can significantly
improve the precision for estimating $\omega$ under the spin relaxation
channel, avoid the exponential loss of the precision for estimating $\gamma$
(estimating $\omega$) under all the decoherence channels (under the spin flip
channel) and recover the Heisenberg scaling for estimating $\omega$ under the
spin dephasing channel.

\begin{acknowledgments}
This work was supported by the MOST of China (Grants No. 2014CB848700), the
NSFC (Grants No. 11322542, No. 11704308, No. 11604261, and No. 11547166), and
the NSFC program for \textquotedblleft Scientific Research
Center\textquotedblright\ (Grant No. U1530401).
\end{acknowledgments}

\appendix{}

\section{Framework of quantum parameter estimation}

\begin{figure}[ptb]
\includegraphics[width=\columnwidth,clip]{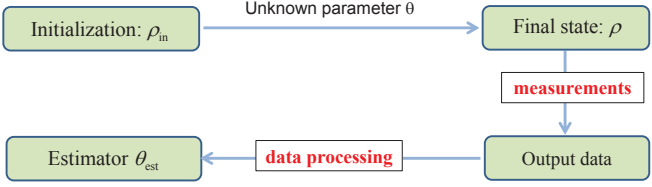}\caption{A typical
framework of quantum parameter estimation.}%
\label{G_FRAMEWORK}%
\end{figure}

Here we provide a self-contained introduction to the typical framework and
important concepts in quantum parameter estimation. A typical (non-adaptive)
parameter estimation protocol using a quantum probe to estimate an unknown,
real parameter $\theta$ consists of three steps (Fig. \ref{G_FRAMEWORK}):

\leftmargini=3mm

\begin{enumerate}
\item The quantum probe starts from an initial state $\rho_{\mathrm{in}}$ and
undergoes certain $\theta$-dependent evolution into a final state $\rho$ that
depends on $\theta$. This step imprints the information about $\theta$ into
the final state $\rho$. The information contained in $\rho$ is quantified by
the quantum Fisher information (QFI)$\ \mathcal{F}$.

\item The quantum probe undergoes a measurement, which produces an outcome
according to certain probability distribution. In this step, the quantum
Fisher information $\mathcal{F}$ contained in $\rho$ is transferred into the
classical information in the measurement outcome. The information contained in
each outcome is quantified by the classical Fisher information (CFI) $F$,
which obeys
\begin{equation}
F\leq\mathcal{F}. \label{FCFQ}%
\end{equation}

\item Steps 1-2 are repeated $\nu$ times and the $\nu$ outcomes are processed
to yield an estimator $\theta_{\mathrm{est}}$ to the unknown parameter
$\theta$. In this step, the total CFI $\nu F$ contained in the $\nu$ outcomes
is converted to the estimation precision, as quantified by the statistical
error of the estimator:
\begin{equation}
\delta\theta\equiv\sqrt{\langle(\theta_{\mathrm{est}}-\theta)^{2}\rangle},
\label{UNCERTAINTY}%
\end{equation}
where $\langle\cdots\rangle$ denotes the average over a lot of estimators
obtained by repeating steps 1-3 many times. For unbiased estimators obeying
$\langle\theta_{\mathrm{est}}\rangle=\theta$, the precision $\delta\theta$ is
fundamentally limited by the inequality%
\begin{equation}
\delta\theta\geq\frac{1}{\sqrt{\nu F}}\Leftrightarrow(\delta\theta)^{-2}%
\leq\nu F, \label{CRB}%
\end{equation}
known as the Cram\'{e}r-Rao bound
\cite{HelstromBook1976,KayBook1993,BraunsteinPRL1994}.
\end{enumerate}

\subsection{Quantum Fisher information}

The amount of information about $\theta$ contained in a general $\theta
$-dependent quantum state $\rho$ is quantified by its QFI
\cite{BraunsteinPRL1994}%
\[
\mathcal{F}\equiv\operatorname*{Tr}\rho L^{2},
\]
where $L$ is the so-called symmetric logarithmic derivative operator: it is an
Hermitian operator defined through \cite{HelstromBook1976}
\[
\partial_{\theta}\rho=\frac{1}{2}(L\rho+\rho L).
\]
The QFI is invariant under any $\theta$-independent unitary transformations.
For a pure state $\rho=|\Phi\rangle\langle\Phi|$, we have $L=2\partial
_{\theta}\rho$ and hence
\begin{equation}
\mathcal{F}=4(\langle\partial_{\theta}\Phi|\partial_{\theta}\Phi
\rangle-|\langle\Phi|\partial_{\theta}\Phi\rangle|^{2}). \label{QFI_PURE}%
\end{equation}
For a general mixed state with the spectral decomposition $\rho=\sum_{n}%
p_{n}|\Phi_{n}\rangle\langle\Phi_{n}|$, its QFI is
\cite{KnyshPRA2011,ZhangPRA2013,LiuPRA2013,JingCTP2014}%
\[
\mathcal{F}[\rho]=\sum_{n}\frac{(\partial_{\theta}p_{n})^{2}}{p_{n}}+\sum
_{n}p_{n}\mathcal{F}[|\Phi_{n}\rangle]-\sum_{m\neq n}\frac{8p_{m}p_{n}}%
{p_{m}+p_{n}}\left\vert \left\langle \Phi_{m}\right\vert \partial_{\theta}%
\Phi_{n}\rangle\right\vert ^{2},
\]
where $\{p_{n}\}$ are \textit{nonzero} eigenvalues of $\rho$, $\{|\Phi
_{n}\rangle\}$ are the corresponding ortho-normalized eigenstates, and
$\mathcal{F}[|\Phi_{n}\rangle]$ is the QFI of the pure state $|\Phi_{n}%
\rangle$ [see Eq. (\ref{QFI_PURE})]. For a two-level system, its density
matrix can always be expressed in terms of the Pauli matrices
$\boldsymbol{\sigma}$ as $\rho=(1/2)(1+\boldsymbol{\sigma}\cdot\mathbf{n})$,
where $\mathbf{n}\equiv\operatorname*{Tr}\boldsymbol{\sigma}\rho$ is the Bloch
vector. The QFI for such a state is
\cite{DittmannJPA1999,ZhongPRA2013,LiPRA2015}
\[
\mathcal{F}=|\partial_{\theta}\mathbf{n}|^{2}+\frac{(\mathbf{n}\cdot
\partial_{\theta}\mathbf{n})^{2}}{1-|\mathbf{n}|^{2}},
\]
where the second term is absent when $|\mathbf{n}|=1$, i.e., when $\rho$ is a
pure state. When $\rho=\rho^{(1)}\otimes\cdots\otimes\rho^{(N)}$ is the direct
product state of $N$ quantum probes, its QFI is additive:\ $\mathcal{F}%
=\sum_{n=1}^{N}\mathcal{F}[\rho^{(n)}]$.

The importance of the QFI for parameter estimation is manifested in the
inequalities Eqs. (\ref{FCFQ}) and (\ref{CRB}). Namely, given $\rho$ and hence
$\mathcal{F}$, the precision of \textit{any} unbiased estimator from $\nu$
repetitions of \textit{any} measurement is limited by the inequality%
\begin{equation}
\delta\theta\geq\frac{1}{\sqrt{\nu\mathcal{F}}}, \label{QCRB_A}%
\end{equation}
known as the quantum Cram\'{e}r-Rao bound
\cite{HelstromBook1976,BraunsteinPRL1994}. Saturating this bound requires
saturating Eqs. (\ref{FCFQ}) and (\ref{CRB}) simultaneously, i.e., using
optimal measurements to convert all the QFI into the CFI and using optimal
unbiased estimators to convert all the CFI into the precision of the estimator.

\subsection{Classical Fisher information and optimal measurements}

A general measurement with discrete outcomes $\{u\}$ is described by the
positive-operator valued measure (POVM) elements $\{M_{u}\}$ satisfying the
completeness relation $\sum_{u}M_{u}^{\dagger}M_{u}=1$. Given a quantum state
$\rho$, it yields an outcome $u$ according to the probability distribution
$P(u|\theta)\equiv\operatorname*{Tr}M_{u}\rho M_{u}^{\dagger}$ that depends on
$\theta$. The amount of information about $\theta$ contained in each outcome
is quantified by the CFI \cite{KayBook1993}:%
\begin{equation}
F\equiv\sum_{u}P(u|\theta)\left(  \frac{\partial\ln P(u|\theta)}%
{\partial\theta}\right)  ^{2}. \label{CFI_DEF}%
\end{equation}
For continuous outcomes, we need only replace $\sum_{u}$ by $\int du$
everywhere. The inequality Eq. (\ref{FCFQ}) expresses the simple fact that no
new information about $\theta$ can be generated in the measurement
process:\ those measurements that convert all (part) of the QFI\ into the CFI
are called optimal (non-optimal). Given $\rho$, the optimal measurement is
\textit{not} unique. The projective measurement on the symmetric logarithmic
derivative operator $L$ has been identified \cite{BraunsteinPRL1994} as an
optimal measurement.

\subsection{Optimal unbiased estimators}

Given the measurement distribution $P(u|\theta)$ and hence the CFI $F$ of each
outcome, the precision $\delta\theta$ of \textit{any} unbiased estimator
$\theta_{\mathrm{est}}(\mathbf{u})$ constructed from the outcomes
$\mathbf{u}\equiv(u_{1},\cdots,u_{\nu})$ of $\nu$ repeated measurements is
limited by the Cram\'{e}r-Rao bound Eq. (\ref{CRB}), which expresses the
simple fact that no new information about $\theta$ can be generated in the
data processing:\ optimal (non-optimal) unbiased estimators convert all (part)
of the CFI into the useful information $(\delta\theta)^{-2}$ quantified by the
precision $\delta\theta$. In the limit of large $\nu$, two kinds of estimators
are known to be unbiased and optimal: the maximum likelihood estimator and the
Bayesian estimator \cite{KayBook1993}, as we introduce now.

Before any measurements, our prior knowledge about the unknown parameter
$\theta$ is quantified by certain probability distribution $P_{0}(\theta)$,
e.g., a $\delta$-like distribution corresponds to knowing $\theta$ exactly, a
flat distribution corresponds to completely no knowledge about $\theta$, while
a Gaussian distribution $P_{0}(\theta)\propto e^{-(\theta-\theta_{0}%
)^{2}/(2\sigma_{0}^{2})}$ corresponds to knowing $\theta$ to be $\theta_{0}$
with a typical uncertainty $\sigma_{0}$.

Upon getting the first outcome $u_{1}$, our knowledge about $\theta$ is
immediately refined from $P_{0}(\theta)$ to%
\[
P_{u_{1}}(\theta)=\frac{P_{0}(\theta)P(u_{1}|\theta)}{\mathcal{N}(u_{1})}%
\]
according to the Bayesian rule \cite{ToussaintRMP2011}, where $\mathcal{N}%
(u_{1})\equiv\int d\theta P_{0}(\theta)P(u_{1}|\theta)$ is a normalization
factor ensuring $P_{u_{1}}(\theta)$ is normalized to unity: $\int P_{u_{1}%
}(\theta)d\theta=1$. Here $P_{u_{1}}(\theta)$ is the \textit{posterior}
probability distribution of $\theta$ conditioned on the outcome of the
measurement being $u_{1}$:\ its parametric dependence on $u_{1}$ means that
different measurement outcomes leads to different refinement of knowledge
about $\theta$.

Upon getting the second outcome $u_{2}$, our knowledge is immediately refined
from $P_{u_{1}}(\theta)$ to
\[
P_{u_{1}u_{2}}(\theta)=\frac{P_{0}(\theta)P(u_{1}|\theta)P(u_{2}|\theta
)}{\mathcal{N}(u_{1},u_{2})},
\]
where $\mathcal{N}(u_{1},u_{2})=\int P_{0}(\theta)P(u_{1}|\theta
)P(u_{2}|\theta)d\theta$ is a normalization factor for the posterior
distribution $P_{u_{1}u_{2}}(\theta)$. If we omit the trivial normalization
factors, then the measurement-induced knowledge refinement becomes
\[
P_{0}(\theta)\overset{u_{1}}{\longrightarrow}P_{0}(\theta)P(u_{1}%
|\theta)\overset{u_{2}}{\longrightarrow}P_{0}(\theta)P(u_{1}|\theta
)P(u_{2}|\theta)\overset{u_{3}}{\longrightarrow}\cdots.
\]

Upon getting $\nu$ outcomes $\mathbf{u}\equiv(u_{1},\cdots,u_{\nu})$, our
knowledge about $\theta$ is quantified by the posterior distribution
\[
P_{\mathbf{u}}(\theta)\sim P_{0}(\theta)P(\mathbf{u}|\theta)
\]
up to a trivial normalization factor, where $P(\mathbf{u}|\theta
)=P(u_{1}|\theta)\cdots P(u_{\nu}|\theta)$ is the probability for getting the
outcome $\mathbf{u}$. The posterior distribution $P_{\mathbf{u}}(\theta)$
completely describe our state of knowledge about $\theta$. Nevertheless,
sometimes a single number, i.e., an unbiased estimator, is required as the
best guess to $\theta$. There are two well-known estimators: the maximum
likelihood estimator \cite{KayBook1993}
\begin{equation}
\theta_{\mathrm{M}}(\mathbf{u})\equiv\arg\max P_{\mathbf{u}}(\theta)
\label{MLE}%
\end{equation}
is the peak position of $P_{\mathbf{u}}(\theta)$ as a function of $\theta$,
while the Bayesian estimator \cite{KayBook1993}
\begin{equation}
\theta_{\mathrm{B}}(\mathbf{u})\equiv\int\theta P_{\mathbf{u}}(\theta
)d\theta\label{BAYESIAN}%
\end{equation}
is the average of $\theta$. For large $\nu$, both estimators are unbiased and
optimal: $\langle\theta_{\alpha}\rangle=\theta$ and $\delta\theta_{\alpha
}=1/\sqrt{\nu F(\theta)}$, where $\alpha=\mathrm{M}$ or $\mathrm{B}$, and
$\langle\cdots\rangle$ denotes the average over a large number of estimators
obtained by repeating the $\nu$-outcome estimation scheme many times and
$\delta\theta_{\alpha}$ is defined as Eq. (\ref{UNCERTAINTY}) or%
\begin{equation}
\delta\theta_{\alpha}=\sqrt{\int[\theta-\theta_{\alpha}(\mathbf{u}%
)]^{2}P_{\mathbf{u}}(\theta)d\theta}. \label{DZ_DEF}%
\end{equation}

\section{Monitoring dephasing of logical qubits}

We consider an odd number $m$ of spin-1/2's and each spin-1/2 is subjected to
an independent spin dephasing channel. The quantum jump operator $S_{z}$ of
the dephasing channel leads to random transitions between the two eigenstates
$|\pm\rangle\equiv(|\uparrow\rangle\pm|\downarrow\rangle)/\sqrt{2}$ of
$\sigma_{x}$ during the evolution. We define the $(m-1)$-component syndrome
operator $\boldsymbol{\Sigma}=(\sigma_{x}^{(1)}\sigma_{x}^{(2)},\cdots
,\sigma_{x}^{(m-1)}\sigma_{x}^{(m)})$, where $\sigma_{x}^{(i)}$ is the Pauli
matrix for the $i$th spin-1/2. The code subspace as spanned by
\cite{DuerPRL2014,LuNC2015}
\begin{align*}
|  &  \Uparrow\rangle\equiv\frac{|+\rangle^{\otimes m}+|-\rangle^{\otimes m}%
}{\sqrt{2}},\\
|  &  \Downarrow\rangle\equiv\frac{|+\rangle^{\otimes m}-|-\rangle^{\otimes
m}}{\sqrt{2}},
\end{align*}
is an eigensubspace of $\boldsymbol{\Sigma}$ with eigenvalue $+1$ for every
component of $\boldsymbol{\Sigma}$. This allows us to monitor the simultaneous
quantum jump of at most $(m-1)/2$ qubits \cite{DuerPRL2014,LuNC2015}, e.g, the
quantum jump of the $i$th spin-1/2 switches the sign of the $(i-1)$th and the
$i$th component of $\mathbf{\Sigma}$, while leaves other components of
$\mathbf{\Sigma}$ intact. When the Hamiltonian commutes with the syndrome
operator, monitoring the quantum jump does not affect the Hamiltonian
evolution. When the Hamiltonian further commutes with all the quantum jump
operators $S_{z}^{(1)},\cdots,S_{z}^{(m)}$, the quantum jump does not affect
the Hamiltonian evolution. In this case, MQT can fully recover the estimation
precision in the noiseless case.

As an example, we consider $H=(\omega/2)\bar{\sigma}_{z}$, where $\bar{\sigma
}_{z}\equiv\sigma_{z}^{\otimes m}$ maps $|\Uparrow\rangle$ to $|\Uparrow
\rangle$ and $|\Downarrow\rangle$ to $-|\Downarrow\rangle$. This Hamiltonian
commutes with the syndrome operator and all the quantum jump operators. The
initial state $|\psi_{0}\rangle=(|\Uparrow\rangle+|\Downarrow\rangle)/\sqrt
{2}$ lies inside the code subspace. In the absence of quantum jumps, the final
state is $|\psi\rangle=(|\Uparrow\rangle+e^{i\omega T}|\Downarrow
\rangle)/\sqrt{2}$, whose QFI $\mathcal{F}_{\omega}[|\psi\rangle]=T^{2}$
attains the Heisenberg scaling with respect to the time cost $T$. By
frequently measuring the syndrome operator, we can track the simultaneous
phase flip of at most $(m-1)/2$ qubits. Since the phase flip commute with $H$,
they always map $|\Uparrow\rangle$ ($|\Downarrow\rangle$) to another
eigenstate of $H$ with the \textit{same} eigenvalue, thus all the detectable
quantum jumps do not cause the loss of the QFI. For example, a quantum jump of
the $k$th qubit at time $t$ maps the $m$-qubit state $(|\Uparrow
\rangle+e^{i\omega t}|\Downarrow\rangle)/\sqrt{2}$ to $(\sigma_{z}%
^{(k)}|\Uparrow\rangle+e^{i\omega t}\sigma_{z}^{(k)}|\Downarrow\rangle
)/\sqrt{2}$, where $\sigma_{z}^{(k)}|\Uparrow\rangle$ ($\sigma_{z}%
^{(k)}|\Downarrow\rangle$) is still an eigenstate of $\bar{\sigma}_{z}$ with
the same eigenvalue $+1$ ($-1$). Afterwards, the evolution under
$H=(\omega/2)\bar{\sigma}_{z}$ leads to the final state $(\sigma_{z}%
^{(k)}|\Uparrow\rangle+e^{i\omega T}\sigma_{z}^{(k)}|\Downarrow\rangle
)/\sqrt{2}$, which contains the same QFI as the jumpless state. Therefore, MQT
allows us to continuously track the quantum trajectory and hence recover the
Heisenberg scaling of the estimation precision without error correction, as
long as in between two round of syndrome measurements, the number of
phase-flipped qubits do not exceed $(m-1)/2$. By contrast, when the quantum
trajectories are not monitored, the QFI\ of the non-selective final state
would decay to zero on time scales $\gg$ single-qubit decoherence time.

The results above are consistent with the conclusion of Refs.
\cite{ZhouNC2018,DemkowiczPRX2017} since the Hamiltonian $H\propto\bar{\sigma
}_{z}=\sigma_{z}^{\otimes m}$ lies outside the Lindblad span of all the
decoherence channels. The results above can also be generalized to $N$ logical
qubits, where\ each logical qubit is composed of $m$ qubits through the
$m$-qubit phase-flip code and is driven by the Hamiltonian $(\omega
/2)\bar{\sigma}_{z}$. Starting from the initial state $(|\Uparrow
\rangle^{\otimes N}+|\Downarrow\rangle^{\otimes N})/\sqrt{2}$, all the
detectable quantum trajectories contain the same QFI\ $N^{2}T^{2}$, which
gives the Heisenberg scaling for the estimation precision about $\omega$.

\section{Estimation precision for spin-flip channel}

\subsection{Ideal continuous monitoring}

For ideal, continuous monitoring of the quantum jump (i.e., spin flip), the
normalized quantum trajectory with $n$ quantum jumps at $t_{1},\cdots,t_{n}$
has a QFI
\[
\mathcal{F}_{\omega}[\rho_{t_{1}\cdots t_{n}}]=\left\{
\begin{array}
[c]{ll}%
4|\rho_{\uparrow\downarrow}|^{2}(T-2t_{1}+2t_{2}-\cdots-2t_{2k+1})^{2} &
(n=2k+1),\\
4|\rho_{\uparrow\downarrow}|^{2}(T+2t_{1}-2t_{2}+\cdots-2t_{2k})^{2} & (n=2k),
\end{array}
\right.
\]
with occurence probability density $p_{t_{1}\cdots t_{n}}=(\gamma
/4)^{n}e^{-\gamma T/4}$. For $n=0$, they reduce to the QFI and the occurence
probability of the jumpless trajectory. The trajectory-averaged QFI follows by
straightforward calculation:%

\[
\mathcal{\bar{F}}_{\omega}=4T^{2}|\rho_{\uparrow\downarrow}|^{2}%
e^{-\frac{\gamma T}{4}}\sum_{k=0}^{\infty}\frac{{(\gamma T/4)}^{2k}}%
{(2k+1)!}\left(  1+\frac{{\gamma T/4}}{2k+3}\right)  .
\]
The sum $\sum_{k=0}^{\infty}(\cdots)$ is equal to $(xe^{x}-\sinh x)/x^{2}$
with $x\equiv\gamma T/4$, so we obtain the $\mathcal{\bar{F}}_{\omega}$ listed
in Table I. Since $p_{t_{1}\cdots t_{n}}$ is independent of $\omega$ and the
timings $\{t_{1},\cdots,t_{n}\}$ of the quantum jumps, Eq. (\ref{CFI}) gives
$F_{\omega}=0$ and%
\begin{equation}
F_{\gamma}=\frac{\left(  \partial_{\gamma}P_{\oslash}\right)  ^{2}}%
{P_{\oslash}}+\sum_{n=1}^{\infty}\frac{(\partial_{\gamma}P_{n})^{2}}{P_{n}%
}=\frac{T}{4\gamma},
\end{equation}
where
\[
P_{n}\equiv\int_{0}^{T}dt_{n}\cdots\int_{0}^{t_{2}}dt_{1}\ p_{t_{1}\cdots
t_{n}}=\frac{(\gamma T/4)^{n}}{n!}e^{-\gamma T/4}%
\]
is the probability for $n$ quantum jumps during $[0,T]$.

\subsection{Realistic QEC-based monitoring}

When we use QEC-based MQT with the syndrome operator $\Sigma\equiv\sigma
_{z}\sigma_{z}^{(a)}$ with eigenvalues $\pm1$ (see Sec. \ref{SEC_QEC}), the
interval $\Delta$ between successive syndrome measurements is finite. The
total evolution interval $[0,T]$ is divided into $N=T/\Delta\gg1$ segments of
length $\Delta$ sandwiched by $N$ syndrome measurements at $t=\Delta
,2\Delta,\cdots,N\Delta$. At $t=0$, the initial state of the spin-1/2 and the
ancilla is an eigenstate of the syndrome operator with eigenvalue $\lambda
_{0}\equiv+1$. For clarity we use $\lambda_{k}$ ($=+1$ or $-1$) to denote the
outcome of the $k$th syndrome measurement, where $k=1,2,\cdots,N$. If the
$k$th syndrome measurement reports a sign switch, i.e., $\lambda_{k}%
=-\lambda_{k-1}$, then the spin-1/2 undergoes an odd number of quantum jumps
during $[(k-1)\Delta,k\Delta]$, otherwise (i.e., $\lambda_{k}=\lambda_{k-1}$)
the spin-1/2 undergoes an even number of quantum jumps during $[(k-1)\Delta
,k\Delta]$. If $\gamma\Delta\ll1$, then the possibility of multiple quantum
jumps between neighboring syndrome measurement is exponentially small, so the
syndrome measurements can well resolve individual quantum jumps. Otherwise,
the syndrome measurements only give little information about the quantum jumps.

First, we derive the evolution of the quantum trajectories between the
$(k-1)$th and the $k$th syndrome measurements. Suppose the previous syndrome
measurements give the outcomes $\lambda_{1},\cdots,\lambda_{k-1}$ and the
un-normalized quantum trajectory $\tilde{\rho}$ immediately after the
$(k-1)$th syndrome measurement is%
\[
\tilde{\rho}\equiv%
\begin{bmatrix}
\tilde{\rho}_{\uparrow\uparrow} & \tilde{\rho}_{\uparrow\downarrow}\\
\tilde{\rho}_{\uparrow\downarrow}^{\ast} & \tilde{\rho}_{\downarrow\downarrow}%
\end{bmatrix}
,
\]
which is an eigenstate of $\Sigma$ with eigenvalue $\lambda_{k-1}$. The
\textit{non-selective} evolution during $[(k-1)\Delta,k\Delta]$ gives the
non-selective final state:
\[
e^{\mathcal{L}\Delta}\tilde{\rho}=\mathcal{M}_{+}\tilde{\rho}+\mathcal{M}%
_{-}\tilde{\rho},
\]
where $\mathcal{L}\rho=-i[H,\rho]+\gamma\mathcal{D}[S_{x}]\rho$ and
$\mathcal{M}_{\pm}$ are superoperators:%
\begin{align*}
\mathcal{M}_{+}\tilde{\rho} &  \equiv e^{-\zeta}%
\begin{bmatrix}
\tilde{\rho}_{\uparrow\uparrow}\cosh\zeta & \tilde{\rho}_{\uparrow\downarrow
}f(\Delta)\\
\tilde{\rho}_{\uparrow\downarrow}^{\ast}f^{\ast}(\Delta) & \tilde{\rho
}_{\downarrow\downarrow}\cosh\zeta
\end{bmatrix}
,\\
\mathcal{M}_{-}\tilde{\rho} &  \equiv e^{-\zeta}%
\begin{bmatrix}
\tilde{\rho}_{\downarrow\downarrow}\sinh\zeta & \tilde{\rho}_{\uparrow
\downarrow}^{\ast}g(\Delta)\\
\tilde{\rho}_{\uparrow\downarrow}g(\Delta) & \tilde{\rho}_{\uparrow\uparrow
}\sinh\zeta
\end{bmatrix}
,
\end{align*}
$\zeta\equiv\gamma\Delta/4$, $f(\Delta)\equiv\cos(w\Delta)-i(\omega
/w)\sin(w\Delta)$, $g(\Delta)\equiv\gamma\sin{(w\Delta)}/(4w)$, and
$w\equiv\sqrt{\omega^{2}-(\gamma/4)^{2}}.$ Since each quantum jump leads to
the exchanges $|\uparrow\rangle\leftrightarrow|\downarrow\rangle$ and hence
$\tilde{\rho}_{\uparrow\uparrow}\leftrightarrow\tilde{\rho}_{\downarrow
\downarrow}$ and $\tilde{\rho}_{\uparrow\downarrow}\leftrightarrow\tilde{\rho
}_{\downarrow\uparrow}$, we identify $\tilde{\rho}_{+}\equiv\mathcal{M}%
_{+}\tilde{\rho}$ ($\tilde{\rho}_{-}\equiv\mathcal{M}_{-}\tilde{\rho}$) as the
un-normalized quantum trajectory containing an even (odd) number of quantum
jumps during $[(k-1)\Delta,k\Delta]$. In other words, the un-normalized
quantum trajectory at $t=k\Delta$ is $\tilde{\rho}_{-}$ if the $k$th syndrome
measurement reports a sign switch, or $\tilde{\rho}_{-}$ otherwise. Their
occurence probabilities are $\operatorname*{Tr}\tilde{\rho}_{\pm}=[(1\pm
e^{-2\zeta})/2]\operatorname*{Tr}\tilde{\rho}$, so the non-selective evolution
is trace-preserving:$\ \operatorname*{Tr}\mathcal{M}_{+}\tilde{\rho
}+\operatorname*{Tr}\mathcal{M}_{-}\tilde{\rho}=\operatorname*{Tr}\tilde{\rho
}$. When $\operatorname*{Tr}\tilde{\rho}=1$, the occurence probabilities of
$\tilde{\rho}_{\pm}$ are $p\equiv(1+e^{-2\zeta})/2$ and $1-p$, respectively.

This single-step evolution can be iterated $N$ times to yield the quantum
trajectories at $t=T$. Specifically, if only the syndrome measurements at
$t=k_{1}\Delta,\cdots,k_{m}\Delta$ report sign switches, then the
un-normalized quantum trajectory at $t=T$ is%
\[
\tilde{\rho}_{k_{1}\cdots k_{m}}=\mathcal{M}_{+}^{N-k_{m}-1}\mathcal{M}%
_{-}\mathcal{M}_{+}^{k_{m}-k_{m-1}-1}\cdots\mathcal{M}_{-}\mathcal{M}%
_{+}^{k_{2}-k_{1}-1}\mathcal{M}_{-}\mathcal{M}_{+}^{k_{1}-1}\rho_{0},
\]
and its occurence probability
\[
P_{k_{1}\cdots k_{m}}\equiv\operatorname*{Tr}\tilde{\rho}_{k_{1}\cdots k_{m}%
}=p^{N-m}(1-p)^{m}%
\]
is independent of $k_{1},\cdots,k_{m}$. Since $p$ is independent of $\omega$,
the CFI about $\omega$ vanishes:\ $F_{\omega}=0$, and the CFI about $\gamma$
follows from Eq. (\ref{CFI}) as%
\[
F_{\gamma}=\sum_{m=0}^{N}\sum_{\{k_{1},\cdots,k_{m}\}}\frac{(\partial_{\gamma
}P_{k_{1}\cdots k_{m}})^{2}}{P_{k_{1}\cdots k_{m}}}=\sum_{m=0}^{N}%
\frac{(\partial_{\gamma}P_{m})^{2}}{P_{m}}=\frac{T}{4\gamma}\frac{4\zeta
}{e^{4\zeta}-1},
\]
where $P_{m}\equiv\sum_{k_{1}<k_{2}<\cdots<k_{m}}P_{k_{1}\cdots k_{m}}%
=C_{N}^{m}p^{N-m}(1-p)^{m}$ is the probability for $m$ quantum jumps during
$[0,T]$ and $C_{N}^{m}\equiv N!/(m!(N-m)!)$ is the binomial coefficient. Next,
we calculate the QFIs in the normalized quantum trajectories: $\rho
_{k_{1}\cdots k_{m}}\equiv\tilde{\rho}_{k_{1}\cdots k_{m}}/P_{k_{1}\cdots
k_{m}}$.

When $\omega=0$, we have $f(\Delta)=\cosh\zeta$, $g(\Delta)=\sinh\zeta$, and
hence $\mathcal{M}_{+}\rho_{0}=p\rho_{0}$ and $\mathcal{M}_{-}\rho
_{0}=(1-p)\sigma_{x}\rho_{0}\sigma_{x}$, so $\rho_{k_{1}\cdots k_{m}}=\rho
_{0}$ (for even $m$) or $\rho_{k_{1}\cdots k_{m}}=\sigma_{x}\rho_{0}\sigma
_{x}$ (for odd $m$) is independent of $\gamma$ and hence give zero trajectory
QFI about $\gamma$. In this case, the total information from MQT is
$\mathbb{F}_{\gamma}=F_{\gamma}$.

When $\omega\neq0$, to obtain explicit analytical expression, we assume
$1/\Delta\gg\omega\gg\gamma$ and $\gamma T\ll1/(\gamma\Delta)$, so that
$f(\Delta)\approx e^{-i\omega\Delta}$, $g(\Delta)\approx\zeta$, and hence
\begin{align*}
\rho_{k_{1}\cdots k_{m}} &  =%
\begin{bmatrix}
\rho_{\downarrow\downarrow} & a\\
a^{\ast} & \rho_{\uparrow\uparrow}%
\end{bmatrix}
\ (\mathrm{for\ }m=2n+1),\\
\rho_{k_{1}\cdots k_{m}} &  =%
\begin{bmatrix}
\rho_{\uparrow\uparrow} & b\\
b^{\ast} & \rho_{\downarrow\downarrow}%
\end{bmatrix}
\ (\mathrm{for}\ m=2n),
\end{align*}
where%
\begin{align*}
a &  \equiv{\frac{\rho_{\uparrow\downarrow}^{\ast}e^{-\gamma T/4}}%
{p^{N-2n-1}(1-p)^{2n+1}}}\zeta^{2n+1}e^{-i\omega\Delta(N+1+2\sum
\limits_{i=1}^{2n+1}(-1)^{i}k_{i})},\\
b &  \equiv\frac{\rho_{\uparrow\downarrow}e^{-\gamma T/4}}{p^{N-2n}(1-p)^{2n}%
}\zeta^{2n}e^{-i\omega\Delta(N+2\sum\limits_{i=1}^{2n}(-1)^{i+1}k_{i})}.
\end{align*}
The trajectory-QFI is
\begin{align*}
\mathcal{F}_{\omega}[\rho_{k_{1}\cdots k_{m}}] &  =4|\rho_{\uparrow\downarrow
}|^{2}\frac{\zeta^{2m}\Delta^{2}e^{-\gamma T/2}}{[p^{N-m}(1-p)^{m}]^{2}}%
\times\\
&  \left\{
\begin{array}
[c]{ll}%
(N-2k_{2n+1}+2k_{2n}-\cdots-2k_{1}+1)^{2} & (m=2n+1),\\
(N-2k_{2n}+2k_{2n-1}-\cdots+2k_{1})^{2} & (m=2n).
\end{array}
\right.
\end{align*}
Averaging the trajectory-QFI over their occurence probabilities gives
\begin{align*}
\mathcal{\bar{F}}_{\omega} &  \approx4|\rho_{\uparrow\downarrow}%
|^{2}e^{-\gamma T/4}\Delta^{2}\sum_{n=0}^{[N/2]}\left(  N\frac{N(N-1)\cdots
(N-2n)}{(2n+1)!}(\zeta e^{-\zeta})^{2n}\right.  \\
&  \left.  +\frac{(N+1)N\cdots(N-2n-1)}{(2n+3)(2n+1)!}(\zeta e^{-\zeta
})^{2n+1}\right)  ,
\end{align*}
where $[x]$ is the largest integer not greater than $x$ and we have used
$p^{N-m}\approx e^{-(N-m)\zeta}$ and $(1-p)^{m}\approx\zeta^{m}$. Performing
the summation gives Eq. (\ref{FW_DELTA}) of the main text.

%

\end{document}